\documentclass[manuscript]{acmart}
\pdfoutput=1
\usepackage{graphicx}
\usepackage{capt-of}

\usepackage[hide]{chato-notes}

\usepackage{amsmath}
\usepackage{comment}
\usepackage{booktabs}
\usepackage{multirow}
\usepackage{paralist}
\usepackage{xspace}
\usepackage{wrapfig}
\definecolor{darkgreen}{rgb}{0,100,0}

\usepackage{multirow}
\usepackage[normalem]{ulem}

\setcopyright{acmcopyright}
\copyrightyear{2022}
\acmYear{2022}
\acmDOI{10.1145/XXXXXXX.XXXXXXX}

\newcommand\refomitted[1]{[\emph{reference omitted for double-blind review}]\xspace}
\newcommand\RisCanvi{{\em RiskEval}\xspace}
\newcommand\Catalonia{{\em Country}\xspace}

\begin{document}

\title[Human Predictions in Algorithm-Supported Recidivism Risk Assessment]{A Comparative User Study of Human Predictions in Algorithm-Supported Recidivism Risk Assessment}

%

\author{Manuel Portela}
\affiliation{%
  \institution{Universitat Pompeu Fabra}
  \streetaddress{Campus Poblenou}
  \city{Barcelona}
  \country{Spain}
}
\email{manuel.portela@upf.edu}
\orcid{1234-5678-9012}

\author{Carlos Castillo}
\affiliation{%
 \institution{ICREA and Universitat Pompeu Fabra}
 \city{Barcelona}
 \country{Spain}}
\email{chato@acm.org}

\author{Songül Tolan}
\affiliation{%
  \institution{European Commission, Joint Research Centre}
  \city{Seville}
  \country{Spain}}
  \email{songul.tolan@ec.europa.eu}

\author{Marzieh Karimi-Haghighi}
\affiliation{%
 \institution{Universitat Pompeu Fabra}
 \city{Barcelona}
 \country{Spain}}
 \email{marzieh.karimihaghighi@upf.edu}

\author{Antonio Andres Pueyo}
\affiliation{%
 \institution{Universitat de Barcelona}
 \city{Barcelona}
 \country{Spain}}
 \email{andrespueyo@ub.edu}
\renewcommand{\shortauthors}{Portela et al.}

\begin{abstract}
In this paper, we study the effects of using an algorithm-based risk assessment instrument to support the prediction of risk of criminal recidivism. 
The instrument we use in our experiments is a machine learning version of \RisCanvi (name changed for double-blind review), which is the main risk assessment instrument used by the Justice Department of \Catalonia (omitted for double-blind review).
The task is to predict whether a person who has been released from prison will commit a new crime, leading to re-incarceration, within the next two years. We measure, among other variables, the accuracy of human predictions with and without algorithmic support.
%
%
This user study is done with (1) {\em general} participants from diverse backgrounds recruited through a crowdsourcing platform, (2) {\em targeted} participants who are students and practitioners of data science, criminology, or social work and professionals who work with \RisCanvi.
Among other findings, we observe that algorithmic support systematically leads to more accurate predictions from all participants, but that statistically significant gains are only seen in the performance of targeted participants with respect to that of crowdsourced participants.
We also run focus groups with participants of the targeted study to interpret the quantitative results, including people who use \RisCanvi in a professional capacity. 
Among other comments, professional participants indicate that they would not foresee using a fully-automated system in criminal risk assessment, but do consider it valuable for training, standardization, and to fine-tune or double-check their predictions on particularly difficult cases.
\end{abstract}


\maketitle

\section{Introduction}
Since the 1970s the use of Risk Assessment Instruments (RAI) in high stakes contexts such as medicine or criminal justice, together with their risks and benefits, have been a subject of debate across various disciplines.
RAIs may increase the accuracy, robustness, and efficiency in decision making \cite{kleinberg2018human}; however, they can also lead to biased decisions and, consequently, to discriminatory outcomes \cite{Angwin2016_machinebias,skeem2016gender}. 

\begin{table}[t]
\hspace{-6mm}\begin{minipage}{0.48\linewidth}
    \centering\includegraphics[width=0.96\linewidth]{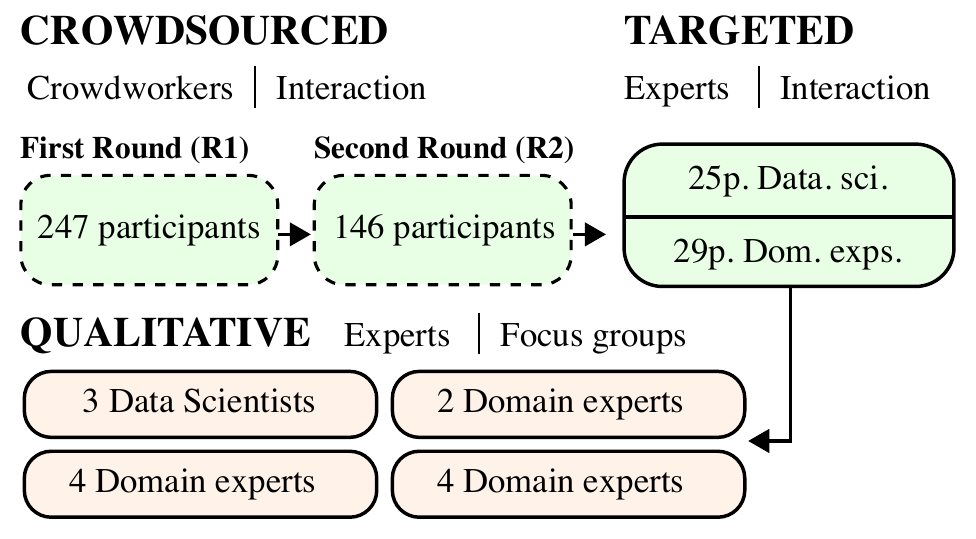}
    
     \captionof{figure}{Sequence of studies and number of participants.}\label{fig:studies}
     \Description{The figure shows the three experimental studies (one targeted and two crowdsourced) and the different focus groups made during our study.}

\end{minipage}%
\hspace{2mm}%
\begin{minipage}{0.51\linewidth}
    \resizebox{1.02\linewidth}{!}{%
     \begin{tabular}{ll|cccc}
    \toprule
      & & \textbf{Crowd. (R1)} & \textbf{Crowd. (R2)} & \textbf{Dom.Exp} & \textbf{Data.Sci} \\
      & & N=247 & N=146 & N=29 & N=25\\
    \midrule
    \textbf{Gender} & Male &  61.0\%  & 54.3\% & 20.7\% & 40.0\% \\
       & Female &  48.4\% & 45.7\% & 79.3\% & 56.0\%   \\
       & Other &  0.6\%  & 0.0\% & 0.0\% & 4.0\% \\
    \midrule
       \textbf{Education}  
        & Secondary & 19.3\% & 15.5\% & 3.5\% & 20.0\% \\
        & Undergraduate & 57.5\% & 68.2\%  & 48.2\% & 56.0\% \\
        & Postgraduate & 22.3\% & 15.5\% & 48.3\% & 24.0\% \\
    \midrule
        \textbf{Age} & 18-25 & 44.3\% & 53.5\%  & 55.2\%  &  76.0\% \\
        & 26-33 & 25.2\% & 24.8\% & 10.3\% & 20.0\% \\
        & 34-45 & 19.6\% & 12.4\% & 13.8\% & 4.0\% \\
        & 45-75 & 10.9\% &  9.2\% & 20.7\% & 0.0\% \\
 
    \midrule
       \textbf{Numeracy}  & 0 (lowest) & 11.2\% & 18.5\% & 17.2\% &  4.0\% \\
    & 1 & 15.0\% & 20.5\% & 24.1\% & 0.0\% \\
    & 2 & 24.0\% & 22.0\% & 6.9\% & 28.0\% \\
    & 3 (highest) & 49.8\% & 39.0\% & 51.7\% & 68.0\% \\
    \bottomrule
  \end{tabular}

    }\caption{Demographics by study group.}\label{tab:demographics}
\end{minipage}
\vspace{-3pt}
 \end{table}

Understanding the performance of a RAI requires looking beyond the statistical properties of a predictive algorithm, and considering the quality and reliability of the decisions made \emph{by humans} using the RAI \cite{Green2021}.
This is because high-stakes decisions are rarely made by algorithms alone, and humans are almost invariably ``in-the-loop,'' i.e., involved to some extent in the decision making process \cite{Binns2021}.
Indeed, the General Data Protection Regulation (GDPR)\footnote{Available online: \url{https://eur-lex.europa.eu/legal-content/EN/TXT/?uri=CELEX\%3A32016R0679}. Accessed Jan 2022.} in Europe gives data subjects the right ``not to be subject to a decision based solely on automated processing'' (Article 22),
and the proposed Artificial Intelligence Act\footnote{Available online: \url{https://eur-lex.europa.eu/legal-content/EN/TXT/?uri=CELEX:52021PC0206}. Accessed Jan 2022.} published in April 2021, considers criminal risk assessment a ``high-risk'' application subject to stringent human oversight.

Our work involves a sequence of studies outlined in Figure~\ref{fig:studies} and described in the next sections. 
We develop and test different user interfaces of a machine learning version of \RisCanvi, the main RAI used by \Catalonia's criminal justice system.\footnote{Name of instrument changed and name of country omitted for double-blind review.} 
%
%
We ask participants to predict the re-incarceration risk based on the same factors used by \RisCanvi, such as criminal history, which empirically affect the recidivism risk of individuals.
Some participants are additionally shown the risk that the RAI predicts using the same factors.
Our primary goal is to assess how the interaction with the studied RAI affects human predictions, their accuracy, and their willingness to rely on a RAI for this task.

As most previous studies on this topic, we partially rely on crowdsourcing \cite{Dressel2018,Grgic-Hlaca2019,Green2020,fogliato2021impact}.
%
Controlled in-lab/survey experiments and crowdsourced experiments have the limitation that participants do not face the real world consequences that professional decisions have on the lives of inmates.
%
In addition, untrained crowdworkers may exhibit different decision making behaviour than trained professionals. 
The former limitation can only be addressed through studies that analyze the real-world adoption of a RAI through observational 
methods \cite{berk2017impact,stevenson2018assessing,stevenson2021algorithmic}. However, these studies usually face the difficulty of isolating the effect of RAI adoption from other changes that co-occur in the study period.
%
%
%
The latter can be addressed in an experimental setting by recruiting professional participants, as we do in this paper.
To the best of our knowledge, most studies focus on crowdsourced participants. This might be the first study that, in addition to a crowdsourced study, runs a targeted study. We recruited students and professionals of data science as well as domain experts (with a background in criminology and social work), including people who work within \Catalonia's criminal justice system and use \RisCanvi in a professional capacity. 
Finally, we conducted a \emph{qualitative study} with small sub-groups of the targeted user study, particularly professionals 
within the Justice Department of \Catalonia, as well as data scientists. Our main contributions are:
\begin{itemize}
\item We confirm previous results that show how accuracy in decision making slightly improves with algorithmic support, how participants adjust their own predictions in the direction of the algorithmic predictions, and how different scales in risk communication yield different levels of accuracy.
\item We describe differences between targeted participants and crowdsourced workers. Despite identical experimental conditions and tasks, we find that the predictions differ between these groups and targeted participants outperform crowdsourced participants in terms of accuracy. 
\item We provide insights into how professionals use RAIs in real-world applications from our focus groups. Our interviewees would not foresee using a fully automated system in criminal risk-assessment, but they see benefits in using algorithmic support for training and standardization, and for fine-tuning and double-checking particularly difficult cases.
\end{itemize}

The remainder of this paper is structured as follows.
First, we give an overview of related work (\S\ref{sec:related-work}).
Next, we describe our approach, including the variables of our study (\S\ref{sec:approach}), as well as the materials and procedures we employ (\S\ref{sec:methods}).
We present the experiment setup for crowdsourced and targeted participants (\S\ref{sec:participants-and-exp-setup}), and the obtained results from both groups (\S\ref{sec:results}). Then we present the results from the focus groups (\S\ref{sec:qualitative-study}).
Finally, we discuss our findings  (\S\ref{sec:discussion}), as well as the limitations of this study and possible directions for future work (\S\ref{sec:limitations}).
\section{Related work}
\label{sec:related-work}

\subsection{Risk Assessment Instruments (RAI) for Criminal Recidivism}
\label{subsec:related-work-observational}

Law enforcement systems increasingly use statistical algorithms, e.g., methods that predict the risk of arrestees to re-offend, to support their decision making \cite{Goel2019,chiusi_automating_2020}. 
RAIs for criminal recidivism risk are in use in various countries including
Austria \cite{rettenberger2010entwicklung},
Canada \cite{kroner2007validity},
Germany \cite{dahle2014development}, 
Spain \cite{andres2018riscanvi},
the U.K. \cite{howard2012construction},
and
the U.S. \cite{desmarais2013risk}.
%
%
There are ethical and legal aspects to consider, as algorithms may exhibit biases, which are sometimes inherited from the data on which they are trained \cite{barocas2016big}.  
However, some argue that RAIs bear the potential for considerable welfare gains \cite{kleinberg2018human}. 
The literature shows that decisions based on RAIs' scores are never made by an algorithm alone. Decisions in criminal justice are made by professionals (e.g., judges or case workers) \cite{bao2021s}, sometimes using RAIs  \cite{stevenson2021algorithmic}.
%
%
Consequently, algorithms aimed at supporting decision processes, especially in high-risk contexts such as criminal justice, cannot be developed without taking into account the influences that institutional, behavioural, and social aspects have on the decisions \cite{Selbst2019}. 
Furthermore, human factors such as biases, preferences and deviating objectives can also influence the effectiveness of algorithm-supported decision making~\cite{jahanbakhsh2020experimental,Mallari2020}.
%
%

%
%
Experienced decision makers may be more inclined to deviate from an algorithmic recommendation, relying more on their own cognitive processes \cite{Green2020}.
Moreover, trained professionals, such as probation officers, may prefer to rely on their own decision and not just on a single numerical RAI prediction. 
Any additional information that they consider may be used as a reason to deviate from what a RAI might recommend for a case \cite{McCallum2017}. 
%
There are other reasons why humans disagree with an algorithmic recommendation.
For instance, the human's objectives might be misaligned with the objective for which the algorithm is optimized \cite{Green2020a}, or the context may create incentives for the human decision maker not to follow the algorithm's recommendation \cite{stevenson2021algorithmic}. 
%
Sometimes humans are unable to evaluate the performance of themselves or the risk assessment, and engage in "disparate interactions" reproducing biased predictions by the algorithm~\cite{Green2019a}.
%
%
Another reason could be algorithm aversion, e.g., human decision makers may discontinue the use of an algorithm after observing a mistake, even if the algorithm is on average more accurate than them \cite{dietvorst2015algorithm,burton2020systematic}.
%
%
In contrast, controlled user studies in criminal risk assessment indicate that crowdsourced participants tend to exhibit \emph{automation bias}, i.e., a tendency to over-rely on the algorithm's prediction \cite{Dressel2018, Bansak2019}.
%

%
%
Effective human-algorithm interaction depends on users' training with the tool, on the experience of the human decision maker with the algorithm, and on the specific professional domain in which the decision is made.
Therefore, some researchers have studied the impact of the adoption of RAIs in criminal justice decision-making in real-world applications \cite{berk2017impact,stevenson2018assessing,stevenson2021algorithmic}.
These observational studies yield valuable insights, but the conditions of adoption as well as the design of the RAI cannot be controlled, making it difficult to isolate the effect of the RAI on the studied outcome.

\subsection{Controlled User Studies and Interaction Design of RAIs}
\label{subsec:related-work-experimental}

Algorithm-supported human decision making has also been studied in controlled experiments \cite{Dressel2018,Green2019,Green2019a,grgic2019human,Lin2020,fogliato2021impact}. 
Among these, an influential study by Dressel and Farid in 2018 \cite{Dressel2018}, showed how crowdsourced users recruited from Amazon Mechanical Turk (AMT) were able to outperform the predictions of COMPAS, a RAI that has been subject to significant scrutiny since the seminal work of Angwin et al.~\cite{Angwin2016_machinebias}.
Follow-up studies criticized Dressel and Farid's study, noting that participants were shown the ground truth of each case (i.e., whether or not the person actually recidivated) immediately after every prediction they make, which does not correspond to how these instruments are used in practice. 
Without this feedback, human predictions that were not supported by algorithms performed worse than the algorithm under analysis \cite{Lin2020}.

The way risk assessments are communicated and integrated in the decision process plays a crucial role in the quality of the predictions.
For instance, criminal forensics clinicians have a preference for (non-numerical) categorical statements (such as ``low risk'' and ``high risk'') over numerical risk levels. However, an experimental survey
showed that a RAI providing numerical information elicits better predictive accuracy than if categorical risk levels are used~\cite{ZoeHilton2008}. One issue with categorical expressions is that professionals tend to disagree about the limits of the categories and how these categories represent different numerical risk estimations \cite{Hilton2015}.
%
%
However, numerical expressions introduce other challenges. For instance, participants in a study perceived violence risk as higher when the risk was presented in a frequency format instead of a percentage format~\cite{Hilton2015}.
Another question is whether numerical risks should be presented on an absolute or a relative scale.
A study with clinicians showed that participants hardly distinguish between absolute probability of violence and comparative risk \cite{ZoeHilton2008}.
%
%
%
Furthermore, besides showing only risk levels, risk assessments could include additional information about the nature of the crime, the factors of the RAI and other factors that may have preventive effects on future re-offense~\cite{Heilbrun1999}.
Complementary and graphical information can improve the understanding of risk evaluations \cite{Hilton2017}.
However, it can also increase the overestimation of risk factors while ignoring other contextual information \cite{Batastini2019}. Nevertheless, the use of different visualization methods is mainly unexplored. 

Given the experience from previous work, we build our user interface to test and measure the performance of participants using different categorical risk levels and numerical expressions for risk, specifically absolute and relative risk scales.
We conduct a recidivism prediction experiment with crowdsourced participants, but also complement it with targeted participants.
One of the main novelties of our study resides in assessing how targeted participants, including domain experts and data scientists, perform differently than crowdsourced participants.
Additionally, focus groups and interviews with professionals provide valuable insights into how RAIs are perceived and used in practice.
\vspace{-2mm}
\section{Approach and research questions}
\label{sec:approach}

This paper takes an experimental approach.
Participants in our experiments are asked to determine the probability that an inmate will be re-arrested, based on a list of criminologically relevant characteristics of the case.
We focus on three main outcome variables (\S\ref{subsec:outcomes}): the accuracy of predictions, the changes that participants make to their predictions when given the chance to revise them after seeing the RAI's recommendation, and their willingness to rely on similar RAIs.
The main independent variables (\S\ref{subsec:conditions}) are the background of the participants, and the type of risk scale used.
Our research questions (\S\ref{subsec:research-questions}) are about the interaction of these variables.

\subsection{Outcome variables}
\label{subsec:outcomes}

\subsubsection{Predictive accuracy}

The performance of predictive tools including RAIs is often evaluated in terms of the extent to which they lead to correct predictions. 
Due to the inherent class imbalance in this domain, as most people do not recidivate, most studies (e.g., experimental \cite{Dressel2018,Harris2015,Green2019}) do not use the metric \emph{accuracy}, which is the probability of issuing a correct prediction.
Instead, it is more common to measure the area under the receiver operating characteristic (\textbf{AUC-ROC} or simply \textbf{AUC}).
%
The AUC can be interpreted as the probability that a randomly drawn recidivist obtains a higher score than a randomly drawn non-recidivist.
%

\subsubsection{Prediction alignment with the RAI}
%
In this work, we observe users' reliance on the algorithmic support system indirectly by looking at changes in their predictions after observing an algorithmic prediction.
We assume that if users change their initial predictions to align them with those of a RAI, they are implicitly signaling more reliance on that RAI than if they would have stuck to their initial prediction.
In general, the extent to which people are willing to trust and rely on a computer system is related to people's engagement and confidence in it \cite{VanMaanen2007,Chancey2017,Lee2004},
and in the case of predictive algorithms, to their perceived and actual accuracy~\cite{Yin2019}.
Different types of information disclosure can elicit different levels of trust and reliance~\cite{Du2019}. 
Performing joint decisions, i.e., being the human in the loop~\cite{De-Arteaga2020}, can increase willingness to rely on a system~\cite{Zhang2020}. 

\subsubsection{Preferred level of automation}\label{sec:level_automation}

The experience of interacting with an algorithm-based RAI may also affect the acceptability of similar algorithms in the future. Algorithm-based RAIs may operate in ways that differ by their \emph{level of automation}~\cite{Cummings2004}.
At the lowest level of automation, the human makes all decisions completely disregarding the RAI;
at the highest level of automation, the RAI makes all decisions without human intervention;
intermediate levels represent various types of automated interventions.
%
%
In general, the level of automation chosen by a user should be proportionate to the performance of the automated system.
%
%
Both \emph{algorithm aversion} \cite{burton2020systematic} or under-reliance, as well as \emph{automation bias} \cite{Mosier1998} or over-reliance, negatively affect the predictive accuracy of users. 
%

\subsection{Participant groups and conditions}
\label{subsec:conditions}

In this section we describe the main independent variables that we tested in the experiments.

\subsubsection{Participant's educational and professional background} 
Most user studies on recidivism risk prediction rely on crowdsourced participants from online platforms.
The background of participants may change the way they interact with a RAI.
Data scientists and statisticians have training on statistics, probability, and predictive instruments.
Domain experts with a background in psychology, criminology, or who work within the prison system, have a deeper knowledge of factors driving criminal recidivism. 
Additionally, domain experts who use RAIs receive training on their usage, and they often have a fair amount of training in applied statistics.

Naturally, in real-world applications case worker decisions are far more consequential than the consequences faced by crowdworkers in their lab-like decision scenarios.
Similar to previous work \cite{Green2019a, Cheng2019, Yu2020, Dressel2018}, we add an incentive (in the form of a bonus payment) for correct predictions in the crowdsourced studies. However, this is to encourage appropriate effort, not to simulate a high-stakes scenario.

We consider three participant groups: (1) crowdsourced workers from unspecified backgrounds, (2) students and practitioners of data science, and (3) students of criminology and people with expertise in the prison system. Recruitment procedures are described in \S\ref{subsec:participant-recruitment}.

\subsubsection{Risk scales}

The literature on risk communication suggests that both numerical and categorical information are useful for different purposes \cite{ZoeHilton2008, Jung2013, McCallum2017, Storey2015}.
Categories alone can be misleading when similar cases are assigned to different categories despite only small differences in their risk \cite{Jung2013}. 
In our research, we initially used only a categorical scale,\footnote{We group probabilities using a five-level, empirically-grounded recommendation developed by the US Department of Justice and the US National Reentry Resource Center~\cite{Hanson2017}.} but then switched to  scales that combine both categorical and numerical values; further, we test two different types of numerical scales.
The first scale is based on the probability of recidivism, which we denote ``absolute scale'' as it expresses a probability in absolute terms.
The second scale we use is based on quantiles of the risk distribution in the data, and we call it the ``relative scale'' since it is relative to the risk levels of other cases in the data. 
We also use five categories, for easier comparison with the absolute scale.
\begin{figure}[t]
\centering
\includegraphics[width=\textwidth]{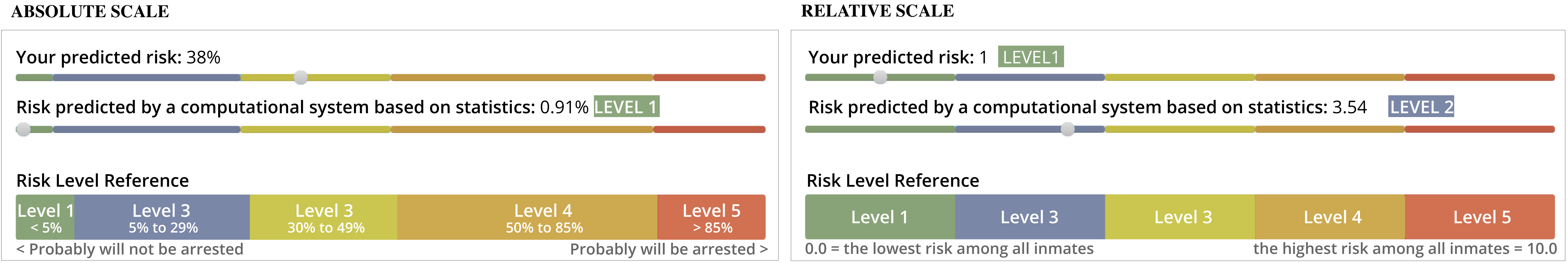}
\captionof{figure}{Risk scales used in our experiments (left: absolute scale, right: relative scale).}
\Description{Shows two different images with the interface where participants can set their prediction. Each image corresponds to different treatment group.\inote{mistake in computatiobal}}
\label{fig:scales}
\vspace{-3pt}
\end{figure}
Both scales are depicted in Figure~\ref{fig:scales}. Other elements in that figure are discussed in the following sections.
    
\subsubsection{Additional variables}
\label{subsubsec:additional-variables}

Many additional variables could have been included but we were mindful of survey length and wanted to minimize survey drop-out. 
We included three additional variables: numeracy, decision-making style, and current emotional state.
Numeracy is the ability to understand and manage numerical expressions.
The decision confidence and the type of information that professionals rely on when using RAIs depends on their numerical proficiency \cite{Scurich2015}.
Ideally, professionals working with RAIs should have a fairly high level of numerical literacy, as interpreting RAIs requires the understanding of probabilities, which is not common knowledge. 
Other factors that have been shown to affect people's decision making behaviour
are their decision-making style and current emotional state~\cite{Beale2008, Lee2012}.

\subsection{Research questions}
\label{subsec:research-questions}

Based on the variables we have presented, we pose the following research questions:

\begin{itemize}
\item[\textbf{RQ1}] \textbf{Under which conditions do participants using a RAI to predict recidivism achieve the highest predictive accuracy?}
\item[\textbf{RQ2}] \textbf{To what extent do participants rely on the RAI to predict recidivism?}
\end{itemize}

\section{Materials and methods}
\label{sec:methods}

In this section, we describe the materials (\S\ref{subsec:materials}) for our user study which consist of a risk prediction instrument based on \RisCanvi (\ref{subsubsec:riscanvi}) and a selection of cases used for assessment (\ref{subsubsec:cases}).
Next, we present a description of the procedure followed by participants (\ref{subsec:procedure}), and the way in which they were recruited (\ref{subsec:participant-recruitment}).

\subsection{Materials}\label{subsec:materials}

\subsubsection{\RisCanvi}\label{subsubsec:riscanvi}

This is one of several risk assessment tools used by the Justice Department of \Catalonia since 
2010 \refomitted{andres2018riscanvi}. 
This tool is applied multiple times during an inmate's time in prison; in most cases, once every six months.

%
%
\RisCanvi consists of 43 items that are completed by professionals based on an inmate's record and suitable interviews. 
%
Then, a team of professionals (with some overlaps with the various interviewers) makes a decision based on the values of the items and the output of \RisCanvi's algorithm.
\RisCanvi's algorithm predicts the risks of four different outcomes: 
committing further violent offenses (violent recidivism),
violence in the prison facilities to other inmates or prison staff,
self-injury,
and breaking of prison permits. 
%
We focus on \emph{violent recidivism}, which is computed based on 23 of the 43 risk factors, including
criminal/penitentiary record,
biographical factors, 
family/social factors,
clinical factors,
and attitude/personality factors. 
%

The original \RisCanvi uses integer coefficients determined by a group of experts; instead, we use a predictor of violent recidivism created using logistic regression that has a better AUC (0.76) than the original \RisCanvi (0.72) 
and is more accurate than models created using other ML methods such as random forests or neural networks~\refomitted{karimi2021enhancing}.
This is done to reduce any effects of potential shortcomings of \RisCanvi originating from using hand-picked integer coefficients, instead using a state-of-the-art predictor based on the same items.


\subsubsection{Cases} \label{subsubsec:cases}

In this study, we use a dataset of cases used in previous work ~\refomitted{karimi2021enhancing}. It consists of the \RisCanvi protocol items for the inmates released between 2010 and 2013, and for which recidivism was evaluated by the Department of Justice of \Catalonia.
Upon recommendation of our Ethics Review Board, we do not show participants the data of any individual inmate, but instead created semi-synthetic cases using a cross-over of cases having similar features and similar risk levels 
(for details, see Supplementary Material~\ref{ann:casesel}). 
%

%
We selected 14 cases which contain a mixture of recidivists and non-recidivists, combining cases in which the majority of humans make correct predictions and cases in which they tend to err, and cases in which the algorithm makes a correct prediction and cases in which it errs.
In our first crowdsourcing experiment (referred to as R1 in the following) we observed that these cases were not representative of the performance of the algorithm on the overall dataset. Hence, for the second crowdsourcing experiment (R2 in the following) we exchanged 2 cases to bring the AUC from 0.61 to 0.75 which is closer to the AUC of the algorithm on the original data (0.76).
Out of the 14 cases, 3 were used as examples during the ``training'' phase of the experiments, while participants were asked to predict recidivism for the remaining 11 cases.
All participants evaluate the same 11 cases, but in randomized order.

\subsection{Procedure} \label{subsec:procedure}

The study obtained the approval of our university's Ethics Review Board in December, 2020. 
All user studies were conducted between December, 2020 and July, 2021, and done remotely due to the pandemic caused by the SARS-COVID-19 virus. 
The survey is designed to be completed in less than 30 minutes and used an interface hosted in our university's server created using standard web technologies (Python and Flask).
The survey is structured as follows:

\subsubsection{Landing page and consent form}
The recruitment (\S\ref{subsec:participant-recruitment}) leads potential participants from different groups to different landing pages, which record which group the participant belongs to.
There, participants learn about the research and we ask for their explicit consent for participating.

\subsubsection{Demographics and additional variables}
Consenting participants are asked three \emph{optional} demographic questions: age (range), gender, and educational level.
Then, three sets of questions are asked to capture the following additional variables (described in \S\ref{subsubsec:additional-variables}):

\noindent- \emph{Numeracy}: We use a test by Lipkus \cite{Lipkus2001}, which has been used in previous work~\cite{Hilton2017}.
It consists of three questions about probabilities, proportions, and percentages, such as 
``If a fair dice is rolled 1,000 times, how many times it will come even (2, 4, or 6)?'' (Answer: 500).
We measure ``numeracy'' as the number of correct answers (0 to 3).

\noindent- \emph{Decision making style}: The General Decision Making  Style (GDMS)~\cite{Scott1995} is a well known survey that    identifies five types of individual decision making style: rational, intuitive, dependent, avoidant, and         spontaneous.  

\noindent- \emph{Current emotional state}: We used a Visual Analogue Scale (VAS) to account for 7 attitudes (happiness, sadness, anger, surprise, anxiety, tranquility, and vigor). This survey has been used in previous work~\cite{Portela2017}.

\subsubsection{Past experience and attitudes towards RAIs}

Participants are asked about their knowledge about and experience with RAIs, as well as what they consider as the three most determining features to predict recidivism, out of the ones used by \RisCanvi.
The final question of this part is about the level of automation they would prefer for determining the risk of recidivism (see Supplementary Material \ref{ann:automationlevel}). 

\subsubsection{Training}

The training part consists of the risk assessment of three cases (two non-recidivists and one recidivist).
The purpose of this part is to prepare participants for the actual evaluations and to calibrate their assessment to a ground truth reference. 
Therefore, unlike the actual risk assessments of the evaluation tasks, participants are shown the ground truth (recidivism or no-recidivism) after each one. 

\subsubsection{Evaluation tasks}

The evaluation tasks are the core part of the study and ask participants to predict the probability of violent recidivism for eleven cases.
Participants see a list of 23 items that are used by \RisCanvi to predict violent recidivism (see Supplementary Material \ref{ann:surveytemplate} for an illustrated reference), and they are asked to select a number, which can be a recidivism probability or a risk level, depending on the condition (see Figure \ref{fig:scales}).
Additionally, they are asked to select from the list of items the three items that they considered most important in their evaluation, and to indicate their confidence with their prediction on a 5-points scale.

Participants in the control group are shown just one screen per case to enter their prediction, while participants in a treatment group are shown a second screen for each case, displaying the algorithm's prediction. 
This second screen also shows participants their initial prediction for comparison, and allows them to optionally change it.
In both screens, participants indicate the confidence in their prediction before continuing. 

\subsubsection{Closing survey}

The experiment ends with a final questionnaire and an evaluation of the entire process.
This questionnaire repeats some of the questions made in the beginning, such as the preferred level of automation, the emotional state, and the three features they consider most important in predicting recidivism. 
Additionally, participants can leave a comment or feedback about the study. 

\subsection{Participant recruitment}\label{subsec:participant-recruitment}

A summary of the participants' demographics is shown in Table \ref{tab:demographics}.
The crowdsourced study consisted of two rounds (\textbf{R1} and \textbf{R2}) for which we recruited participants via Prolific.\footnote{Prolific is a crowdsourcing platform specialized in supporting scientific research. It is available at: https://www.prolific.co/} We selected residents of \Catalonia, 
between 18 to 75 years old, and with more than 75\% of successful completion of other studies in the platform.
Participants were payed a platform-standard rate of 7.5 GBP\footnote{Prolific is a UK-based company that uses British pounds as main currency. We follow their advice for average payment per hour.} per hour for participating in the survey.
They took an average of 20$-$25 minutes to complete the survey. Additionally, we offered a bonus payment of 1 GBP to those who achieved an AUC greater than 0.7. 
This is common practice and incentivizes conscientious completion of the survey (see, e.g., \cite{Green2019a, Cheng2019, Yu2020, Dressel2018}).

For the targeted studies, participants were recruited through students' mailing lists from two universities in \Catalonia, as well as social media groups of professionals of data science in countries having the same official language as \Catalonia. 
Additionally, we invited professionals from the Justice Department of \Catalonia to participate; the invitation to participate was done by their Department of Research and Training. 
The number of participants in previous crowdsourced user studies is usually a few hundred: 103 in \cite{Grgic-Hlaca2019}, 202 in \cite{Cheng2019}, 400 in \cite{Lin2020}, 462 in \cite{Dressel2018} and 600 in \cite{Green2019a}. 
In line with the previous studies, we had 449 participants in total (393 crowdsouced and 54 targeted).

\section{Participants and experimental setup}
\label{sec:participants-and-exp-setup}

\subsection{Crowdsourced: First Round (R1)}
\label{sec:crowdsourced-study}
In the \textbf{first round} (R1) we compared two experimental groups. The  \textbf{treatment} group was  shown the machine prediction and the \textbf{control} group was not.
In treatment group \textbf{G1} machine predictions are shown only as \emph{categorical} information, while in \textbf{G2} machine predictions are shown as \emph{categorical and numerical} information.
In this round, 247 participants completed the evaluation: 48 in the control group, 100 in treatment group G1, and 99 in treatment group G2. 
%
Additionally, 74 participants were excluded, either because they did not complete the survey or did not evaluate all of the eleven cases, or finished the experiment either too fast (less than five minutes) or too slowly (more than one hour).

As described in \S\ref{subsubsec:cases}, we used in R1 a set of cases for which the AUC of the machine predictions was 0.61. 
To bring this more in line with the observed AUC in the entire dataset (0.76), we exchanged two cases for the second round (R2), and the AUC measured on the new set of cases became 0.75.

\subsection{Crowdsourced: Second Round (R2)}
\label{sec:crowdsourced-study-second-round}
\smallskip\noindent In the \textbf{second round} (R2) we compared two experimental groups, where the \textbf{treatment} group was shown the machine prediction and the \textbf{control} group was not.
In treatment group \textbf{G1} machine predictions are shown on an \emph{absolute scale} as categorical and numerical information, while in \textbf{G2} machine predictions are shown on a \emph{relative scale} as categorical and numerical information.

In this round, 146 participants completed the evaluation: 17 in the control group, 66 in treatment group G1, and 63 in treatment group G2. 
%
Additionally, 137 participants were excluded for the same reasons as in R1.

\subsection{Targeted Study}
\label{sec:targeted-study}

The targeted study seeks to establish the effect (if any) of the participant's background when interacting with the RAI.
We used the same experimental setup and treatment groups from crowdsourcing (R2).
Due to the limited number of participants, we considered as a baseline the control group of R2.

We considered both students and professionals with a background either in data science, or in a field relevant to the prison system and the application of \RisCanvi, such as psychology, criminology, or social work.
For data science, we recruited 14 students at the undergraduate and graduate level, and 11 professionals.
For a domain-specific background, we recruited 4 students at the graduate level (Master in Criminology students), and 25 professionals. 
A summary of all experimental groups is shown in Table~\ref{tab:experimentalgroups}.
\section{Results}
\label{sec:results}
\subsection{Outcome variables}
\subsubsection{Predictive accuracy}

\begin{figure}[!tbp]
  \centering
  \begin{minipage}[b]{0.49\textwidth}
    \includegraphics[width=\textwidth]{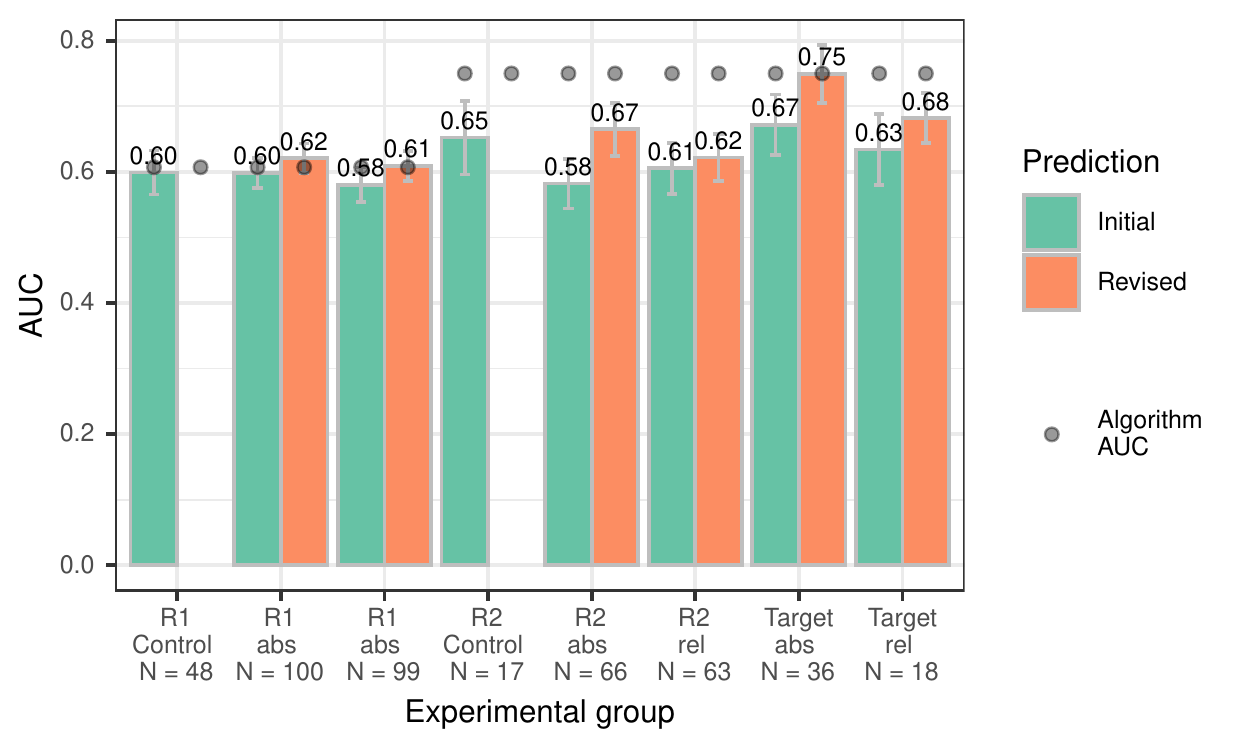}
  \caption{Average AUC with 95\% confidence interval by group. See Table \ref{tab:results1} in Supplementary Material~\ref{ann:accuracy-per-subgroup} for details.}
  \Description{Shows a bar plot comparing each of the experimental groups, comparing their AUC-ROC before presenting the algorithm prediction and after.}
  \label{fig:auc_group}
  \end{minipage}
  \hfill
  \begin{minipage}[b]{0.49\textwidth}
    \includegraphics[width=\textwidth]{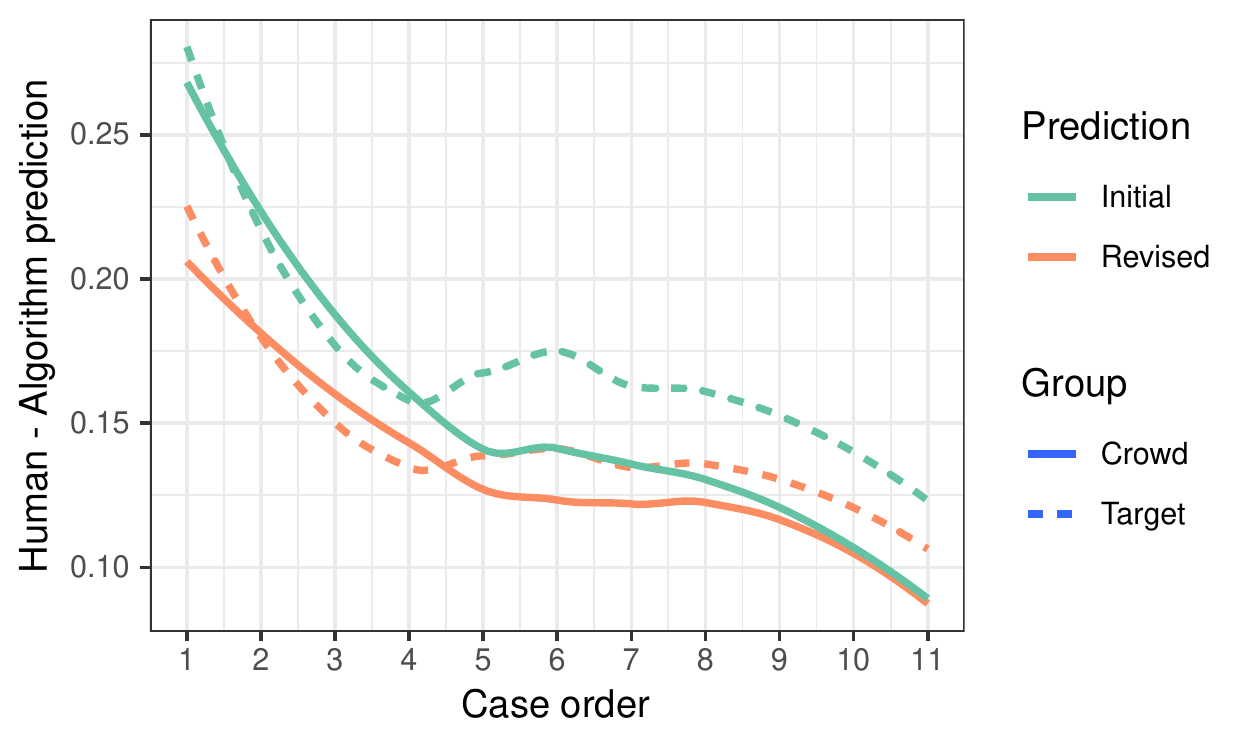}
  \caption{Average difference between human and algorithm prediction by case order, absolute scale.}\label{fig:learning}
  \Description{Shows a line graph where the initial prediciton made by humans is compared with the afterwards, lines are also split between crowdsourced and targeted studies.}
  \end{minipage}
\end{figure}
Figure~\ref{fig:auc_group} shows the average AUC and corresponding confidence intervals\footnote{Confidence intervals in Figure~\ref{fig:auc_group} are computed under normal assumption but we test statistical significance of the differences using a permutation t-test (see Table \ref{tab:ttests})} for each experimental group. This data can also be found in Supplementary Material~\ref{ann:accuracy-per-subgroup}.
%
For R1 we observe no difference in the \emph{initial} predictions across control and treatment groups, which have AUC between 0.58 and 0.61. However, for R2 we find a significant difference (p<0.1) with a higher AUC for the control group (0.65) than for the \emph{initial} prediction of treatment group G1 (0.58) with the absolute scale. 

Despite the small number of participants in the targeted group, we observe important differences compared to the previous groups. The predictive accuracy of the initial prediction is higher (+0.02 to +0.09 AUC points) than any crowdsourced group. For the targeted group G1 (absolute scale) this difference is significant at 
$p<0.1$ against R2's G2 and even at $p<0.05$ against the initial predictions of the other crowdsourced groups (see Supplementary Material~\ref{ann:accuracy-per-subgroup}).
Participants from a data science background and domain experts have similar initial AUCs. 

The resulting AUC is comparable to previous forensic studies that achieved AUCs on average in the range of 0.65$-$0.78 using non-algorithmic RAIs \cite{desmarais2016performance, douglas2003evaluation, singh2011comparative}.

\subsubsection{Prediction changes due to the RAI}

The observed probability of a participant changing a prediction after observing the machine prediction is 
20\% (19\% in G1 and 21\% in G2).
Crowdsourced participants revised their prediction in about 26\% of the cases they examined (27\% in G1, 25\% in G2).
Domain experts revised their prediction in 37\% of the cases, and data scientists in only 13\% of the cases.

Figure~\ref{fig:learning} shows the average difference in risk predictions by human and algorithm for each case. 
Targeted participants started with predictions that were in general as high as those of the crowdsourced groups and equally far from those of the RAI.
As they progress through the evaluation tasks, participants tend to align more and more their predictions with the machine predictions (even in their initial predictions) and the difference between initial and revised predictions diminishes.
For the last three cases, the crowdsourced group's predictions, which are already close to the machine predictions, do not change, while the targeted group maintains a larger difference between initial and revised predictions.

By comparing the average AUC in Figure~\ref{fig:auc_group} and Table~\ref{tab:results1}, we can see that revised predictions from crowdsourced groups tend to be more accurate than their initial ones in terms of AUC.
This difference is significant for R1's G2 ($p<0.1$) and for R2's G1 ($p<0.05$), as shown in Table \ref{tab:ttests} (in Supplementary Material~\ref{ann:accuracy-per-subgroup}). For the targeted groups, we see an improvement in the range from +0.01 to +0.09 AUC points on average.
In almost all cases, revised predictions by the treatment groups are more accurate than those of the control groups. However, few of these differences are statistically significant. 

In general, the average self-reported confidence is in the range 3.5-3.9 out of 5.0 (1.0=least confident, 5.0=most confident), and basically does not change from the initial to the final prediction.
The self-reported confidence of crowdworkers is, by a small but statistically significant margin ($p<0.001$), higher than the one of targeted participants (see Supplementary Material~\ref{ann:self-reported-confidence}).

\subsubsection{Preferred level of automation}

\begin{figure}[!tbp]
  \centering
  \begin{minipage}[b]{0.45\textwidth}
    \includegraphics[width=\textwidth]{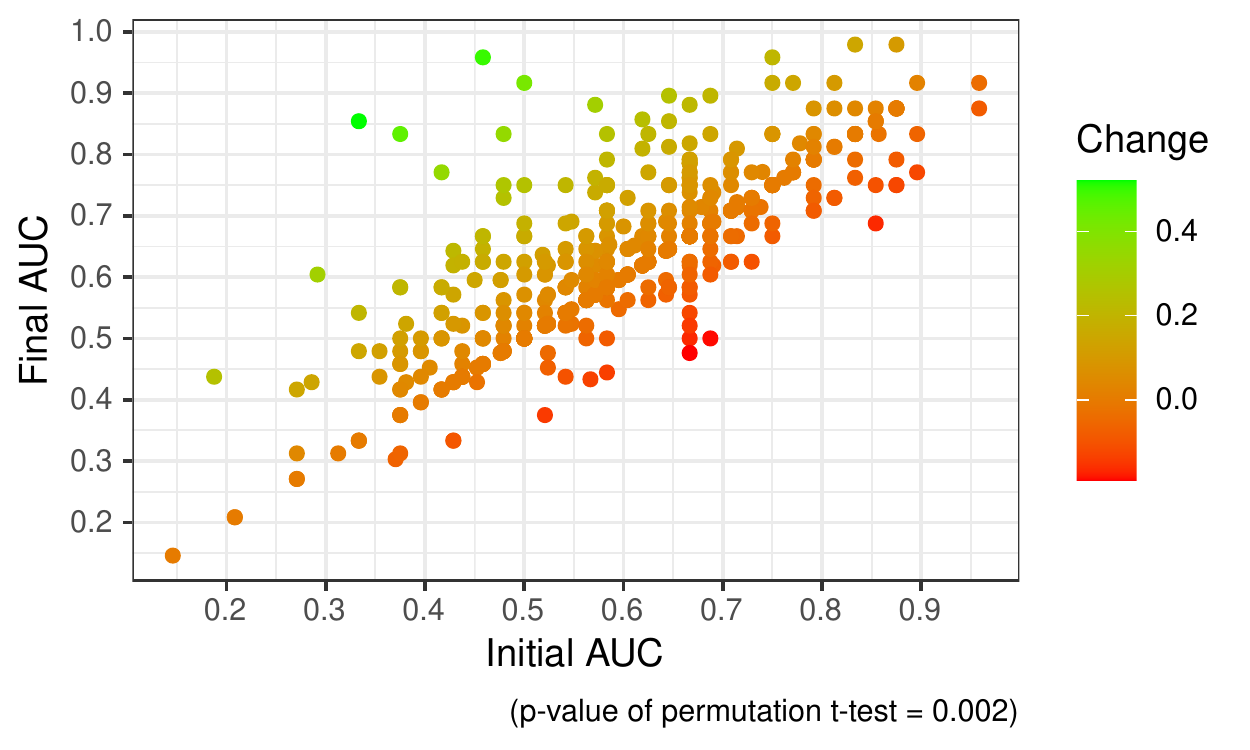}
  \caption{AUC of participant predictions before and after algorithmic support for participants who received algorithmic support (excludes control group).}
  \label{fig:accuracydist}
  \Description{Shows a scatterplot with individual accuracy, both for initial and final prediction (after algorithm suggestion).}
  \end{minipage}
  \hfill
  \begin{minipage}[b]{0.45\textwidth}
    \includegraphics[width=\textwidth]{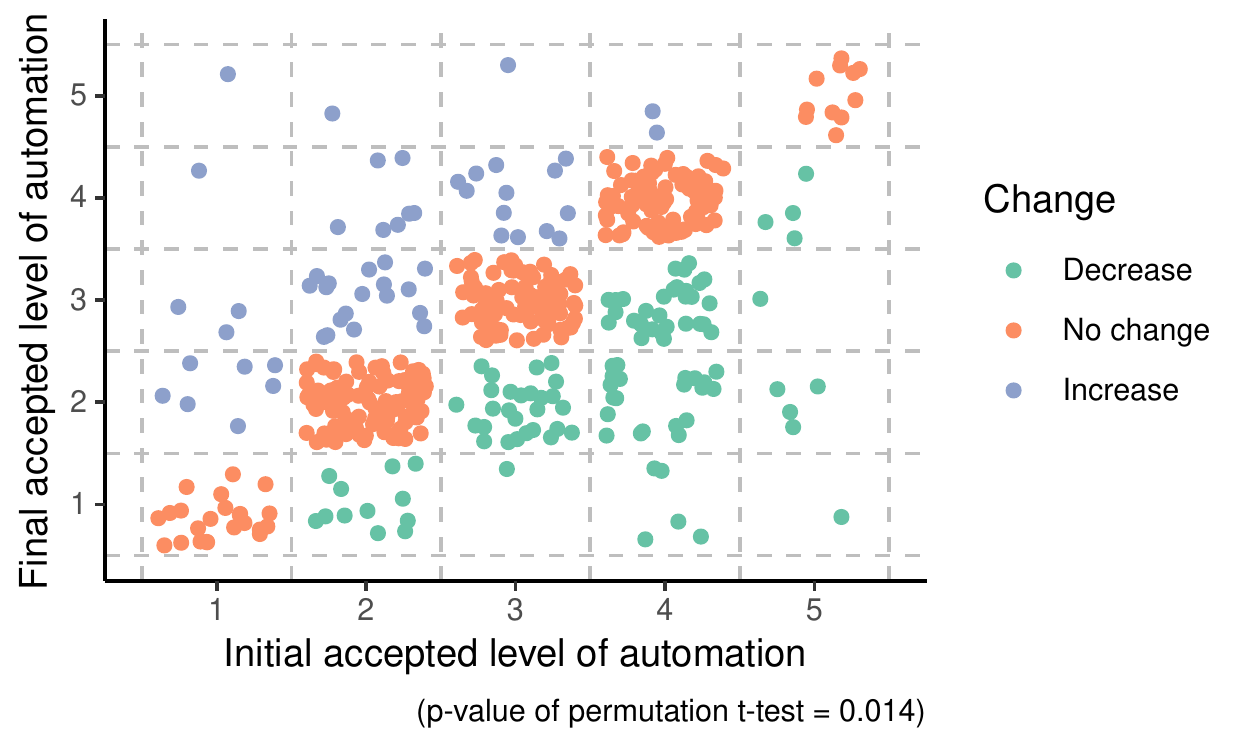}
  \caption{Distribution of answers about level of automation for participants who received algorithmic support (excludes control group).} 
  \label{fig:reliance_nocontrol}
  \Description{Shows a scatterplot divided in sections regarding their answer for the level of automation survey. In the plot, the initial values and the final values are compared.}
  \end{minipage}
  \vspace{-3pt}
\end{figure}

As shown in Figure~\ref{fig:reliance_nocontrol}, most participants prefer an intermediate level of automation, between levels 2-4 on a scale of 5 levels. While data scientists had an initial level of acceptance with a broader range (levels 1-4), domain experts limited their answers to a more narrow set of choices in intermediate levels of acceptance (levels 2-3).
The same figure also shows that most of the treated groups reduce their preferred level of automation after the experiment, meaning they prefer more expert involvement and less reliance on machine predictions.

On average, however, the desired level of acceptance for targeted groups concentrated in the middle-low part of the scale: 32\% of the data scientists and 48\% of the domain experts selected level 3 (``the computational system suggests one option, but the expert can provide an alternative option'').
Level 2 (``the computational system suggests some options, but it is the expert who defines the risk level''), was the option selected by 36\% of the surveyed data scientists and 38\% of the domain experts. 
Details can be found in Supplementary Material~\ref{ann:preferred-level-of-automation}, and the description of the automation levels can be found in Supplementary Material~\ref{ann:automationlevel}.

\subsection{Independent variables}

\subsubsection{Self-reported importance of risk items}
\label{importanceofriskitems}
Having asked to select the top 3 items (risk factors) that participants considered in their risk prediction, we find that crowdsourced and targeted participants tend to select the same 10-11 (out of 23) items as more important than the rest.
However, among these top 10 items we find that domain experts prefer dynamic factors (i.e., factors that can change), such as 
\textit{``limited response to psychological treatment''}, 
while data scientists and crowdsourced participants refer more often than domain experts to static factors (i.e., factors that cannot change), such as \textit{``history of violence''} 
%
(details are in Figure \ref{fig:items_important} and Table \ref{tab:top5_items} in  Supplementary Material~\ref{ann:items-considered-most-important}).

\subsubsection{Numerical information (R1)}

According to Figure~\ref{fig:auc_group}, adding numerical values to the categorical scale does not change the AUC. In G1, where only categorical information is shown, the AUC of revised predictions is slightly higher than the revised predictions in G2, where categorical and numerical values are shown: 0.62 against 0.61 AUC.

\subsubsection{Risk scales (R2 and Targeted)} \label{subsubsec:scales}

The results of R2 show that for the initial prediction, the absolute scale (G1) leads to slightly lower AUC compared to the relative scale (G2) (0.58 against 0.61 AUC).
However, with algorithmic support, the absolute scale leads to higher AUC than the relative scale (0.67 against 0.62 AUC).
Neither of these differences is statistically significant.
Additionally, the average AUC of the R2 control group (0.65) is fairly high, and the only higher AUC observation is in the revised predictions using the absolute scale (0.67).
The revised and some initial predictions of the targeted participants using the absolute scale 
significantly outperform all the R1 groups, as well as the R2 groups ($p<0.05$, see Table~\ref{tab:ttests} in Supplementary Material~\ref{ann:accuracy-per-subgroup}).

\subsubsection{Additional variables}

With respect to \emph{numeracy}, over 60\% of the crowdsourced participants answered correctly 2 or 3 out of the 3 test questions.
The targeted group had more respondents answering all 3 numeracy questions correctly than crowdworkers, as shown in Table~\ref{tab:demographics}: 96\% of data scientists obtained results in the highest scores (68\% in the top score), while only 59\% of domain experts obtained similar results (52\% in the top score).
We find no correlation between participants' numeracy and their accuracy. 
The correlation between \emph{decision making style} and \emph{emotional state} with accuracy is not significant either (results are in Supplementary Material~\ref{ann:gdms-vas}). 

\section{Qualitative study}
\label{sec:qualitative-study}

The last study is a qualitative study using focus groups, i.e., groups of participants having a focused discussion on a particular topic \cite{morgan1998focus}.
The focus groups help us interpret the quantitative results from the targeted study, by listening to and learning from participants' experiences and opinions.

\subsection{Participants and procedure}

Participants (9 women, 4 men) were recruited from the \emph{targeted} experiment, and due to their busy schedules, divided into four groups (FG1-FG4) as follows:
\textbf{FG1} (N=3) data scientists; \textbf{FG2} (N=4) domain experts, students from criminology in undergraduate and master levels; \textbf{FG3} (N=2) and \textbf{FG4} (N=4) domain experts working with the Department of Justice, most of them psychologists.

While we did not want to give too much structure to the conversation, to try to uncover new perspectives that we had not thought about, we did prepare a series of questions to stimulate a discussion (available in Supplementary Material \ref{ann:focusgroup}). 
The questions address participants' experience with algorithmic predictions and RAIs, their opinion about different scales and categorical/numerical presentation, their understanding of risk factors, and their desired level of automation.
Each session lasted between 60 and 90 minutes and was held online. Following the protocol approved by our Ethics Review Board, participants were asked for their consent to participate and to have the meeting recorded and transcribed. 
The language of the focus group was the local language spoken in \Catalonia;  
the quotes we present in the next section were taken from our transcriptions and paraphrased in English.

\subsection{Findings}

We focus on our research questions, but note that there were many other insightful comments during the focus groups.

\subsubsection{Professional background}

All participants were aware that some demographics are over/under represented among prison populations, and thus expected that a RAI trained on such data may lead to discriminatory outcomes.
However, the way in which data science participants approached risk prediction was to a large extent based on considering a set of ``anchors'' or prototypes \cite[p. 13]{Scurich2012b}:
``I think about a maximum and a minimum risk. The minimum would be like a person who stole for the first time [...] 
the maximum would be a killer'' (FG1.1).
%
In general, data scientists did not question the presented case characteristics, but domain experts did.
Participants in FG3 and FG4 indicated that the risk items, which in \RisCanvi only have three levels (Yes/Maybe/No), do not accurately represent the reality of inmates
and they were missing the ability to explore or negotiate the risk items during the case evaluations.
Furthermore, they indicated that, during the assignment of levels to risk factors, they sometimes ``compensate'' higher values in one item with lower values in other items, 
%
such that the final risk score matches what they would consider appropriate for the evaluated person.
One participant (FG4.1) 
said that personal biases may also affect the coding of items, as some professionals adopt a more punitive approach, while others take a more protective or rehabilitative approach.
Other domain experts agreed with this perspective.
%
Therefore, most professionals expressed the need for teams reviews and validation mechanisms for risk factor codings. 

Among domain experts, the psychologists we interviewed were the most concerned about the evidence they collect and the representation of the actual risk. 
To them, RAIs are tools that add objectivity to their case reports,
but their focus was on \emph{how} to present evidence to judges, since these might discard professional reports in favor of the RAI's outcome. 
%
%
%
Overall, for domain experts RAIs such as \RisCanvi should be used by a group of experienced evaluators checking one another, and not by one professional alone. 

\subsubsection{Interpreting numbers}

All participants had some training in statistics, and stated that they understand numerical expressions well. 
Generally, participants preferred a relative scale (e.g., 3.7/10.0) over an absolute scale (e.g., 37\%). 
 
It is noteworthy how domain experts interpret probabilities.
 First, 
 extremely low risks were considered unlikely in practice, since almost everyone can commit a crime at some point.
 Second, all interviewed domain experts stated that recidivism risk cannot be eliminated but it could be reduced to an acceptably low level (e.g., reducing the risk from 37\% to 20\%).
 This emphasis on risk reduction is in line with the ``interventions over predictions'' debate in the literature~\cite{barabas2018interventions}.
%
Third, domain experts consider a recidivism risk of above 30\% as high, and a reason for concern. A risk above 50\% 
was considered difficult -but not impossible- to reduce by treatment/interventions.
Overall, domain experts thought of different ranges on the risk spectrum along which inmates are placed.
Data scientists, too, considered different risk ranges, and for some of them even a 50\% recidivism risk was not considered ``high.''

\subsubsection{Interaction with machine predictions and calibration}

Many participants admitted that they went quickly, and without giving it much thought, through the first few evaluations. 
However, they also noticed that they slowed down to rethink when they felt contested by the algorithm, i.e., when their risk assessment was far from the algorithm's prediction. 
Data scientists indicated that they reacted to such differences by simply adjusting the risk to half-way between their initial prediction and the one of the algorithm. 
Domain experts indicated to react similarly in some cases, 
but they also stressed that they kept their initial prediction when they felt confident about it.

Some of the domain experts 
believed that they were interacting with exactly the same \RisCanvi algorithm they use, despite a clear indication in the introduction of the study that this was another algorithm.
We believe their experience with the real \RisCanvi 
affected their disposition to rely on the machine predictions we presented.

\subsubsection{Preferred level of automation}
 
Overall, domain experts and data scientists differed in the level of automation they would prefer, with data scientists being more open to automation.  
For instance, participant FG1.2 
believed that an algorithm could improve enough to make almost-autonomous decisions ``in the future.'' 
This participant considered the errors that could be made by the algorithm were ``acceptable.'' 
In contrast, e.g., FG1.3 
was sceptical about using an algorithm for automated decision-making 
because of the impossibility to solve all algorithm-specific errors. 

All participants agreed that algorithmic support is useful in many instances, e.g., to 
contrast their own predictions, to give them a chance to rethink them, or to provide reassurance about them.
Domain experts also considered them useful to train new case workers in writing evaluations.
In that regard, participants from FG1 and FG2, expressed that the ``objectivity''  of the algorithm could help reduce the effect of the ``emotional'' response to the evidence by the professional who is evaluating. 

%

Participants also acknowledged the risk of ``relying too much'' on the algorithm, leading to reduced professional responsibility: ``The decision you make is yours, it is yours with your biases and everything, which also brings experience because it sometimes helps you to be aware and review your own prejudices'' (FG2.1). 
%
%
Another drawback of using a RAI noted by participants was the concern that it may reproduce potentially outdated societal prejudices.
To address this concern, domain experts expected frequent updates to the algorithms.

\section{Discussion}
\label{sec:discussion}

\textbf{RQ1: Under which conditions do participants using a RAI to predict recidivism achieve the highest predictive accuracy?} 

Overall, our findings suggest that human decision makers achieve higher accuracy for their risk-assessment when they are supported by an algorithm.
%
Almost all treatment groups achieve a higher AUC than their corresponding control group after the treatment, although some of these differences are not statistically significant, particularly in the case of crowdsourced participants (Figures \ref{fig:auc_group} and \ref{fig:accuracydist}). 
The algorithm also influences human predictions for each decision and over time, as shown in Figure \ref{fig:learning}. This further suggests that algorithmic support establishes reference points to human predictions.
The lower accuracy of the initial predictions of treatment group participants compared to control group participants is noteworthy.
One possible explanation for this is that treated participants put less effort in their initial predictions in anticipation of algorithmic support and a potential opportunity to revise their initial prediction. 

The finding that targeted participants (domain experts and data scientists) outperform crowdsourced participants contradicts the idea that crowdsourced participants are comparable to domain experts or professionals when testing RAIs. This highlights the importance of testing RAIs in the context of professional knowledge, training and usage. 

Finally, using an absolute rather than a relative scale leads to more accurate predictions. The focus group further confirmed the preference of professionals for the absolute scale as the one closer to the real application. 
Our findings agree with Hilton et al. \cite{ZoeHilton2008}, who found that risk categories are generally hard to agree upon across professions and individuals,
and also with Hanson et al. \cite{Hanson2017}, who found that categories can be effective following a common agreement in correspondence to ranges of the absolute probability of recidivism.
Thus, further studies should focus on the underlying support of numerical information in helping ground categorical distinctions for predictive risk assessment. 
\noindent\textbf{RQ2: To what extent do participants rely on the RAI to predict recidivism?}

In line with previous studies (e.g., \cite{Tan2018a}) humans and algorithms tend to agree on very low and very high risk cases (see Supplementary Material~\ref{ann:casesel}, particularly Figure \ref{fig:casespredictions}),
but there are cases that are difficult to predict for humans, for algorithms, or for both.
A promising next step would be to identify cases that are clearly difficult for the machine, and or are potentially difficult to humans. In these cases one could more safely defer to humans, or ask them to invest more time in a specific evaluation, improving efficiency in the design of human-algorithm decision processes. 

Our findings show that 
participants prefer a partially automated assistance with a large degree of human discretion.
In addition, all experimental groups tend to downgrade the acceptable level of automation after the experiment (see Figure~\ref{fig:reliance_nocontrol}).
Explanations for this could be that the differences between human and machine predictions caused the participants to realize strong human oversight was more necessary than what they initially thought.
%


Finally, the focus group discussions revealed that professionals' reliance in an algorithm could be increased when the algorithm providers ensure good prediction performance and frequent system updates corresponding to new societal and institutional developments. This suggests that \RisCanvi and possibly other RAIs are elements of negotiation that should be taken with care and without assuming its outcome as objective, and that need frequent updates and audits.

\vspace{-4mm}
\section{Limitations and Future Work}
\label{sec:limitations}

This paper has to be seen in light of some limitations. 
First, 
the dataset used for training the algorithm has some drawbacks. 
It has only about 597 cases, 
which may affect the algorithm's accuracy; however, we note that its AUC-ROC is in line with that of most recidivism prediction tools.
We also note that in this dataset the ground-truth label is \textit{re-arrest} and not \textit{re-offense}, and  re-arrest is not necessarily a good proxy for re-offense and further exhibits racial and geographical disparities \cite{fogliato2021validity}.
Since the focus of this study is the assessment of user behaviour (not the algorithm), we do not expect these drawbacks to notably affect our main results.
Second, in line with previous work, this study focuses on accuracy as a measure of algorithmic performance. 
However, decision support algorithms can be evaluated in many different ways \cite{Sambasivan2021}. 
%
Third, Figure \ref{fig:learning} shows that participants are still calibrating their predictions after the training phase as they progress through the evaluation tasks, suggesting that the initial training phase may have been to short. 
The impact should be limited as the majority of the cases are evaluated after this learning curve has flattened.

The generalization of this work to other contexts is restricted by other factors. As usual in experimental user studies, 
the crowdsourced participants are not representative of the overall population.
Table \ref{tab:demographics} shows that most have university-level education and good numeracy. 
Further, we only recruited participants in a single country. Thus, the pool of users might not exhibit a large cultural diversity, a factor that could bias outcomes~\cite{Beale2008, Lee2012}.
However, we also remark that crime and recidivism is different in different criminal systems and jurisdictions, and hence RAIs should be evaluated with careful attention to their context \cite{Selbst2019}. 
Sample size may be another limitation. While the size of our participant pool in the crowdsourced study (N=247, N=146) is in line with previous work, the number of participants in the targeted study (N=54) is relatively small.
%
Despite these limitations in sample size, our results suggest consistent and in some cases statistically significant differences in the outcomes between crowdsourced and targeted participants.

Future research is needed to explore the reasons and conditions of these differences. This is particularly important in the public sector, where there is a lack of evidence on how algorithms affect public policies \cite{Zuiderwijk2021}. 
There is a clear need to pay attention to the usage contexts and the ways in which RAIs are deployed, to reduce the risks of automation and understand better in which conditions the assistance of an algorithm can be most helpful.

\begin{acks}
This work has been partially supported by the HUMAINT programme (Human Behaviour and Machine Intelligence), Centre for Advanced Studies, Joint Research Centre, European Commission through the Expect contract CT-EX2019D347180.
We thanks the collaboration from Prof. Antonio Andres Pueyo from Universitat de Barcelona (Spain), the Justice Department of Catalonia and the CEJFE.
We thanks Emilia Gomez from the Joint Research Centre, European Commission; Alexandra Chouldechova and Riccardo Fogliato from Carnegie Mellon University, for their invaluable contributions and support. 
The research project was approved by the University's ethical committee (CIREP) certifying that complies with the data protection legal framework, namely, with European General Data Protection Regulation (EU)
2016/679 -GDPR- and Spanish Organic Law 3/2018, of December 5th, on Protection of Personal Data and digital rights guarantee -LOPDGDD-.
\end{acks}

\bibliographystyle{ACM-Reference-Format}
\bibliography{main}


\begin{thebibliography}{70}


\ifx \showCODEN    \undefined \def \showCODEN     #1{\unskip}     \fi
\ifx \showDOI      \undefined \def \showDOI       #1{#1}\fi
\ifx \showISBNx    \undefined \def \showISBNx     #1{\unskip}     \fi
\ifx \showISBNxiii \undefined \def \showISBNxiii  #1{\unskip}     \fi
\ifx \showISSN     \undefined \def \showISSN      #1{\unskip}     \fi
\ifx \showLCCN     \undefined \def \showLCCN      #1{\unskip}     \fi
\ifx \shownote     \undefined \def \shownote      #1{#1}          \fi
\ifx \showarticletitle \undefined \def \showarticletitle #1{#1}   \fi
\ifx \showURL      \undefined \def \showURL       {\relax}        \fi
\providecommand\bibfield[2]{#2}
\providecommand\bibinfo[2]{#2}
\providecommand\natexlab[1]{#1}
\providecommand\showeprint[2][]{arXiv:#2}

\bibitem[\protect\citeauthoryear{Andr{\'e}s-Pueyo, Arbach-Lucioni, and
  Redondo}{Andr{\'e}s-Pueyo et~al\mbox{.}}{2018}]%
        {andres2018riscanvi}
\bibfield{author}{\bibinfo{person}{Antonio Andr{\'e}s-Pueyo},
  \bibinfo{person}{Karin Arbach-Lucioni}, {and} \bibinfo{person}{Santiago
  Redondo}.} \bibinfo{year}{2018}\natexlab{}.
\newblock \showarticletitle{The RisCanvi: a new tool for assessing risk for
  violence in prison and recidivism}.
\newblock \bibinfo{journal}{\emph{Recidivism Risk Assessment: A Handbook for
  Practitioners}} (\bibinfo{year}{2018}), \bibinfo{pages}{255--268}.
\newblock


\bibitem[\protect\citeauthoryear{Angwin, Larson, Mattu, and Kirchner}{Angwin
  et~al\mbox{.}}{2016}]%
        {Angwin2016_machinebias}
\bibfield{author}{\bibinfo{person}{Julia Angwin}, \bibinfo{person}{Jeff
  Larson}, \bibinfo{person}{Surya Mattu}, {and} \bibinfo{person}{Lauren
  Kirchner}.} \bibinfo{year}{2016}\natexlab{}.
\newblock \showarticletitle{Machine Bias: There’s Software Used across the
  Country to Predict Future Criminals and It’s Biased against Blacks}.
\newblock \bibinfo{journal}{\emph{ProPublica}} (\bibinfo{year}{2016}).
\newblock
\newblock
\shownote{Retrieved from
  https://www.propublica.org/article/machine-bias-risk-assessmentsin-criminal-sentencing}.


\bibitem[\protect\citeauthoryear{Bansak}{Bansak}{2019}]%
        {Bansak2019}
\bibfield{author}{\bibinfo{person}{Kirk Bansak}.}
  \bibinfo{year}{2019}\natexlab{}.
\newblock \showarticletitle{{Can nonexperts really emulate statistical learning
  methods? A comment on "the accuracy, fairness, and limits of predicting
  recidivism"}}.
\newblock \bibinfo{journal}{\emph{Political Analysis}} (\bibinfo{year}{2019}),
  \bibinfo{pages}{370--380}.
\newblock
\showISSN{14764989}
\urldef\tempurl%
\url{https://doi.org/10.1017/pan.2018.55}
\showDOI{\tempurl}


\bibitem[\protect\citeauthoryear{Bao, Zhou, Zottola, Brubach, Desmarais,
  Horowitz, Lum, and Venkatasubramanian}{Bao et~al\mbox{.}}{2021}]%
        {bao2021s}
\bibfield{author}{\bibinfo{person}{Michelle Bao}, \bibinfo{person}{Angela
  Zhou}, \bibinfo{person}{Samantha Zottola}, \bibinfo{person}{Brian Brubach},
  \bibinfo{person}{Sarah Desmarais}, \bibinfo{person}{Aaron Horowitz},
  \bibinfo{person}{Kristian Lum}, {and} \bibinfo{person}{Suresh
  Venkatasubramanian}.} \bibinfo{year}{2021}\natexlab{}.
\newblock \showarticletitle{It's COMPASlicated: The Messy Relationship between
  RAI Datasets and Algorithmic Fairness Benchmarks}.
\newblock \bibinfo{journal}{\emph{arXiv preprint arXiv:2106.05498}}
  (\bibinfo{year}{2021}).
\newblock


\bibitem[\protect\citeauthoryear{Barabas, Virza, Dinakar, Ito, and
  Zittrain}{Barabas et~al\mbox{.}}{2018}]%
        {barabas2018interventions}
\bibfield{author}{\bibinfo{person}{Chelsea Barabas}, \bibinfo{person}{Madars
  Virza}, \bibinfo{person}{Karthik Dinakar}, \bibinfo{person}{Joichi Ito},
  {and} \bibinfo{person}{Jonathan Zittrain}.} \bibinfo{year}{2018}\natexlab{}.
\newblock \showarticletitle{Interventions over predictions: Reframing the
  ethical debate for actuarial risk assessment}. In
  \bibinfo{booktitle}{\emph{Conference on Fairness, Accountability and
  Transparency}}. PMLR, \bibinfo{pages}{62--76}.
\newblock


\bibitem[\protect\citeauthoryear{Barocas and Selbst}{Barocas and
  Selbst}{2016}]%
        {barocas2016big}
\bibfield{author}{\bibinfo{person}{Solon Barocas} {and}
  \bibinfo{person}{Andrew~D Selbst}.} \bibinfo{year}{2016}\natexlab{}.
\newblock \showarticletitle{Big data's disparate impact}.
\newblock \bibinfo{journal}{\emph{California Law Review}}
  \bibinfo{volume}{104} (\bibinfo{year}{2016}), \bibinfo{pages}{671}.
\newblock
\urldef\tempurl%
\url{https://heinonline.org/HOL/P?h=hein.journals/calr104&i=695}
\showURL{%
\tempurl}


\bibitem[\protect\citeauthoryear{Batastini, Hoeffner, Vitacco, Morgan, Coaker,
  and Lester}{Batastini et~al\mbox{.}}{2019}]%
        {Batastini2019}
\bibfield{author}{\bibinfo{person}{Ashley~B. Batastini},
  \bibinfo{person}{Camden~E. Hoeffner}, \bibinfo{person}{Michael~J. Vitacco},
  \bibinfo{person}{Robert~D. Morgan}, \bibinfo{person}{Lauren~C. Coaker}, {and}
  \bibinfo{person}{Michael~E. Lester}.} \bibinfo{year}{2019}\natexlab{}.
\newblock \showarticletitle{{Does the Format of the Message Affect What Is
  Heard? A Two-Part Study on the Communication of Violence Risk Assessment
  Data}}.
\newblock \bibinfo{journal}{\emph{Journal of Forensic Psychology Research and
  Practice}} \bibinfo{volume}{19}, \bibinfo{number}{1} (\bibinfo{year}{2019}),
  \bibinfo{pages}{44--71}.
\newblock
\showISSN{24732842}
\urldef\tempurl%
\url{https://doi.org/10.1080/24732850.2018.1538474}
\showDOI{\tempurl}


\bibitem[\protect\citeauthoryear{Beale and Peter}{Beale and Peter}{2008}]%
        {Beale2008}
\bibfield{author}{\bibinfo{person}{Russell Beale} {and}
  \bibinfo{person}{Christian Peter}.} \bibinfo{year}{2008}\natexlab{}.
\newblock \showarticletitle{{The role of affect and emotion in HCI}}.
\newblock \bibinfo{journal}{\emph{Lecture Notes in Computer Science (including
  subseries Lecture Notes in Artificial Intelligence and Lecture Notes in
  Bioinformatics)}}  \bibinfo{volume}{4868 LNCS} (\bibinfo{year}{2008}),
  \bibinfo{pages}{1--11}.
\newblock
\showISBNx{3540850988}
\showISSN{03029743}
\urldef\tempurl%
\url{https://doi.org/10.1007/978-3-540-85099-1_1}
\showDOI{\tempurl}


\bibitem[\protect\citeauthoryear{Berk}{Berk}{2017}]%
        {berk2017impact}
\bibfield{author}{\bibinfo{person}{Richard Berk}.}
  \bibinfo{year}{2017}\natexlab{}.
\newblock \showarticletitle{An impact assessment of machine learning risk
  forecasts on parole board decisions and recidivism}.
\newblock \bibinfo{journal}{\emph{Journal of Experimental Criminology}}
  \bibinfo{volume}{13}, \bibinfo{number}{2} (\bibinfo{year}{2017}),
  \bibinfo{pages}{193--216}.
\newblock


\bibitem[\protect\citeauthoryear{Binns and Veale}{Binns and Veale}{2021}]%
        {Binns2021}
\bibfield{author}{\bibinfo{person}{Reuben Binns} {and} \bibinfo{person}{Michael
  Veale}.} \bibinfo{year}{2021}\natexlab{}.
\newblock \showarticletitle{{Is That Your Final Decision? Multi-Stage
  Profiling, Selective Effects, and Article 22 of the GDPR}}.
\newblock \bibinfo{journal}{\emph{International Data Privacy Law}}
  \bibinfo{volume}{00}, \bibinfo{number}{0} (\bibinfo{year}{2021}),
  \bibinfo{pages}{1--14}.
\newblock


\bibitem[\protect\citeauthoryear{Burton, Stein, and Jensen}{Burton
  et~al\mbox{.}}{2020}]%
        {burton2020systematic}
\bibfield{author}{\bibinfo{person}{Jason~W Burton}, \bibinfo{person}{Mari-Klara
  Stein}, {and} \bibinfo{person}{Tina~Blegind Jensen}.}
  \bibinfo{year}{2020}\natexlab{}.
\newblock \showarticletitle{A systematic review of algorithm aversion in
  augmented decision making}.
\newblock \bibinfo{journal}{\emph{Journal of Behavioral Decision Making}}
  \bibinfo{volume}{33}, \bibinfo{number}{2} (\bibinfo{year}{2020}),
  \bibinfo{pages}{220--239}.
\newblock


\bibitem[\protect\citeauthoryear{Chancey, Bliss, Yamani, and Handley}{Chancey
  et~al\mbox{.}}{2017}]%
        {Chancey2017}
\bibfield{author}{\bibinfo{person}{Eric~T. Chancey}, \bibinfo{person}{James~P.
  Bliss}, \bibinfo{person}{Yusuke Yamani}, {and} \bibinfo{person}{Holly~A.H.
  Handley}.} \bibinfo{year}{2017}\natexlab{}.
\newblock \showarticletitle{{Trust and the Compliance-Reliance Paradigm: The
  Effects of Risk, Error Bias, and Reliability on Trust and Dependence}}.
\newblock \bibinfo{journal}{\emph{Human Factors}} \bibinfo{volume}{59},
  \bibinfo{number}{3} (\bibinfo{year}{2017}), \bibinfo{pages}{333--345}.
\newblock
\showISSN{15478181}
\urldef\tempurl%
\url{https://doi.org/10.1177/0018720816682648}
\showDOI{\tempurl}


\bibitem[\protect\citeauthoryear{Cheng, Wang, Zhang, O'Connell, Gray, Harper,
  and Zhu}{Cheng et~al\mbox{.}}{2019}]%
        {Cheng2019}
\bibfield{author}{\bibinfo{person}{Hao-Fei Cheng}, \bibinfo{person}{Ruotong
  Wang}, \bibinfo{person}{Zheng Zhang}, \bibinfo{person}{Fiona O'Connell},
  \bibinfo{person}{Terrance Gray}, \bibinfo{person}{F.~Maxwell Harper}, {and}
  \bibinfo{person}{Haiyi Zhu}.} \bibinfo{year}{2019}\natexlab{}.
\newblock \showarticletitle{{Explaining Decision-Making Algorithms through
  UI}}. In \bibinfo{booktitle}{\emph{Proceedings of the 2019 CHI Conference on
  Human Factors in Computing Systems - CHI '19}}. \bibinfo{publisher}{ACM
  Press}, \bibinfo{address}{New York, New York, USA}, \bibinfo{pages}{1--12}.
\newblock
\showISBNx{9781450359702}
\urldef\tempurl%
\url{https://doi.org/10.1145/3290605.3300789}
\showDOI{\tempurl}


\bibitem[\protect\citeauthoryear{Chiusi, Fischer, Kayser-Bril, and
  Spielkamp}{Chiusi et~al\mbox{.}}{2020}]%
        {chiusi_automating_2020}
\bibfield{author}{\bibinfo{person}{Fabio Chiusi}, \bibinfo{person}{Sarah
  Fischer}, \bibinfo{person}{Nicolas Kayser-Bril}, {and}
  \bibinfo{person}{Matthias Spielkamp}.} \bibinfo{year}{2020}\natexlab{}.
\newblock \bibinfo{booktitle}{\emph{Automating {Society} {Report} 2020}}.
\newblock \bibinfo{type}{{T}echnical {R}eport}.
  \bibinfo{institution}{AlgorithmWatch}.
\newblock
\urldef\tempurl%
\url{https://automatingsociety.algorithmwatch.org}
\showURL{%
\tempurl}


\bibitem[\protect\citeauthoryear{Cummings}{Cummings}{2004}]%
        {Cummings2004}
\bibfield{author}{\bibinfo{person}{M.~L. Cummings}.}
  \bibinfo{year}{2004}\natexlab{}.
\newblock \showarticletitle{{Automation bias in intelligent time critical
  decision support systems}}.
\newblock \bibinfo{journal}{\emph{Collection of Technical Papers - AIAA 1st
  Intelligent Systems Technical Conference}}  \bibinfo{volume}{2}
  (\bibinfo{year}{2004}), \bibinfo{pages}{557--562}.
\newblock
\showISBNx{156347719X}
\urldef\tempurl%
\url{https://doi.org/10.4324/9781315095080-17}
\showDOI{\tempurl}


\bibitem[\protect\citeauthoryear{Dahle, Biedermann, Lehmann, and
  Gallasch-Nemitz}{Dahle et~al\mbox{.}}{2014}]%
        {dahle2014development}
\bibfield{author}{\bibinfo{person}{Klaus-Peter Dahle},
  \bibinfo{person}{J{\"u}rgen Biedermann}, \bibinfo{person}{Robert~JB Lehmann},
  {and} \bibinfo{person}{Franziska Gallasch-Nemitz}.}
  \bibinfo{year}{2014}\natexlab{}.
\newblock \showarticletitle{The development of the Crime Scene Behavior Risk
  measure for sexual offense recidivism.}
\newblock \bibinfo{journal}{\emph{Law and human behavior}}
  \bibinfo{volume}{38}, \bibinfo{number}{6} (\bibinfo{year}{2014}),
  \bibinfo{pages}{569}.
\newblock


\bibitem[\protect\citeauthoryear{De-Arteaga, Fogliato, and
  Chouldechova}{De-Arteaga et~al\mbox{.}}{2020}]%
        {De-Arteaga2020}
\bibfield{author}{\bibinfo{person}{Maria De-Arteaga}, \bibinfo{person}{Riccardo
  Fogliato}, {and} \bibinfo{person}{Alexandra Chouldechova}.}
  \bibinfo{year}{2020}\natexlab{}.
\newblock \showarticletitle{{A Case for Humans-in-the-Loop: Decisions in the
  Presence of Erroneous Algorithmic Scores}}. In
  \bibinfo{booktitle}{\emph{Proceedings of the 2020 CHI Conference on Human
  Factors in Computing Systems}}. \bibinfo{publisher}{ACM},
  \bibinfo{address}{New York, NY, USA}, \bibinfo{pages}{1--12}.
\newblock
\showISBNx{9781450367080}
\urldef\tempurl%
\url{https://doi.org/10.1145/3313831.3376638}
\showDOI{\tempurl}
\showeprint[arxiv]{2002.08035}


\bibitem[\protect\citeauthoryear{Desmarais and Singh}{Desmarais and
  Singh}{2013}]%
        {desmarais2013risk}
\bibfield{author}{\bibinfo{person}{Sarah Desmarais} {and} \bibinfo{person}{Jay
  Singh}.} \bibinfo{year}{2013}\natexlab{}.
\newblock \showarticletitle{Risk assessment instruments validated and
  implemented in correctional settings in the United States}.
\newblock \bibinfo{journal}{\emph{Lexington, KY: Council of State Governments}}
  (\bibinfo{year}{2013}).
\newblock


\bibitem[\protect\citeauthoryear{Desmarais, Johnson, and Singh}{Desmarais
  et~al\mbox{.}}{2016}]%
        {desmarais2016performance}
\bibfield{author}{\bibinfo{person}{Sarah~L Desmarais},
  \bibinfo{person}{Kiersten~L Johnson}, {and} \bibinfo{person}{Jay~P Singh}.}
  \bibinfo{year}{2016}\natexlab{}.
\newblock \showarticletitle{Performance of recidivism risk assessment
  instruments in US correctional settings.}
\newblock \bibinfo{journal}{\emph{Psychological Services}}
  \bibinfo{volume}{13}, \bibinfo{number}{3} (\bibinfo{year}{2016}),
  \bibinfo{pages}{206}.
\newblock


\bibitem[\protect\citeauthoryear{Dietvorst, Simmons, and Massey}{Dietvorst
  et~al\mbox{.}}{2015}]%
        {dietvorst2015algorithm}
\bibfield{author}{\bibinfo{person}{Berkeley~J Dietvorst},
  \bibinfo{person}{Joseph~P Simmons}, {and} \bibinfo{person}{Cade Massey}.}
  \bibinfo{year}{2015}\natexlab{}.
\newblock \showarticletitle{Algorithm aversion: People erroneously avoid
  algorithms after seeing them err.}
\newblock \bibinfo{journal}{\emph{Journal of Experimental Psychology: General}}
  \bibinfo{volume}{144}, \bibinfo{number}{1} (\bibinfo{year}{2015}),
  \bibinfo{pages}{114}.
\newblock


\bibitem[\protect\citeauthoryear{Douglas, Ogloff, and Hart}{Douglas
  et~al\mbox{.}}{2003}]%
        {douglas2003evaluation}
\bibfield{author}{\bibinfo{person}{Kevin~S Douglas}, \bibinfo{person}{James~RP
  Ogloff}, {and} \bibinfo{person}{Stephen~D Hart}.}
  \bibinfo{year}{2003}\natexlab{}.
\newblock \showarticletitle{Evaluation of a model of violence risk assessment
  among forensic psychiatric patients}.
\newblock \bibinfo{journal}{\emph{Psychiatric Services}} \bibinfo{volume}{54},
  \bibinfo{number}{10} (\bibinfo{year}{2003}), \bibinfo{pages}{1372--1379}.
\newblock


\bibitem[\protect\citeauthoryear{Dressel and Farid}{Dressel and Farid}{2018}]%
        {Dressel2018}
\bibfield{author}{\bibinfo{person}{Julia Dressel} {and} \bibinfo{person}{Hany
  Farid}.} \bibinfo{year}{2018}\natexlab{}.
\newblock \showarticletitle{{The accuracy, fairness, and limits of predicting
  recidivism}}.
\newblock \bibinfo{journal}{\emph{Science Advances}} \bibinfo{volume}{4},
  \bibinfo{number}{1} (\bibinfo{year}{2018}), \bibinfo{pages}{1--6}.
\newblock
\showISSN{23752548}
\urldef\tempurl%
\url{https://doi.org/10.1126/sciadv.aao5580}
\showDOI{\tempurl}


\bibitem[\protect\citeauthoryear{Du, Huang, and Yang}{Du et~al\mbox{.}}{2019}]%
        {Du2019}
\bibfield{author}{\bibinfo{person}{Na Du}, \bibinfo{person}{Kevin~Y. Huang},
  {and} \bibinfo{person}{X.~Jessie Yang}.} \bibinfo{year}{2019}\natexlab{}.
\newblock \showarticletitle{{Not All Information Is Equal: Effects of
  Disclosing Different Types of Likelihood Information on Trust, Compliance and
  Reliance, and Task Performance in Human-Automation Teaming}}.
\newblock \bibinfo{journal}{\emph{Human Factors}} (\bibinfo{year}{2019}).
\newblock
\showISSN{15478181}
\urldef\tempurl%
\url{https://doi.org/10.1177/0018720819862916}
\showDOI{\tempurl}


\bibitem[\protect\citeauthoryear{Fogliato, Chouldechova, and Lipton}{Fogliato
  et~al\mbox{.}}{2021a}]%
        {fogliato2021impact}
\bibfield{author}{\bibinfo{person}{Riccardo Fogliato},
  \bibinfo{person}{Alexandra Chouldechova}, {and} \bibinfo{person}{Zachary
  Lipton}.} \bibinfo{year}{2021}\natexlab{a}.
\newblock \showarticletitle{The Impact of Algorithmic Risk Assessments on Human
  Predictions and its Analysis via Crowdsourcing Studies}.
\newblock \bibinfo{journal}{\emph{arXiv preprint arXiv:2109.01443}}
  (\bibinfo{year}{2021}).
\newblock


\bibitem[\protect\citeauthoryear{Fogliato, Xiang, Lipton, Nagin, and
  Chouldechova}{Fogliato et~al\mbox{.}}{2021b}]%
        {fogliato2021validity}
\bibfield{author}{\bibinfo{person}{Riccardo Fogliato}, \bibinfo{person}{Alice
  Xiang}, \bibinfo{person}{Zachary Lipton}, \bibinfo{person}{Daniel Nagin},
  {and} \bibinfo{person}{Alexandra Chouldechova}.}
  \bibinfo{year}{2021}\natexlab{b}.
\newblock \showarticletitle{On the Validity of Arrest as a Proxy for Offense:
  Race and the Likelihood of Arrest for Violent Crimes}.
\newblock \bibinfo{journal}{\emph{arXiv preprint arXiv:2105.04953}}
  (\bibinfo{year}{2021}).
\newblock


\bibitem[\protect\citeauthoryear{Goel, Shroff, Skeem, and Slobogin}{Goel
  et~al\mbox{.}}{2019}]%
        {Goel2019}
\bibfield{author}{\bibinfo{person}{Sharad Goel}, \bibinfo{person}{Ravi Shroff},
  \bibinfo{person}{Jennifer~L. Skeem}, {and} \bibinfo{person}{Christopher
  Slobogin}.} \bibinfo{year}{2019}\natexlab{}.
\newblock \showarticletitle{{The Accuracy, Equity, and Jurisprudence of
  Criminal Risk Assessment}}.
\newblock \bibinfo{journal}{\emph{SSRN Electronic Journal}}
  (\bibinfo{year}{2019}), \bibinfo{pages}{1--21}.
\newblock
\urldef\tempurl%
\url{https://doi.org/10.2139/ssrn.3306723}
\showDOI{\tempurl}


\bibitem[\protect\citeauthoryear{Green}{Green}{2020}]%
        {Green2020a}
\bibfield{author}{\bibinfo{person}{Ben Green}.}
  \bibinfo{year}{2020}\natexlab{}.
\newblock \showarticletitle{{The false promise of risk assessments: Epistemic
  reform and the limits of fairness}}.
\newblock \bibinfo{journal}{\emph{FAT* 2020 - Proceedings of the 2020
  Conference on Fairness, Accountability, and Transparency}}
  (\bibinfo{year}{2020}), \bibinfo{pages}{594--606}.
\newblock
\showISBNx{9781450369367}
\urldef\tempurl%
\url{https://doi.org/10.1145/3351095.3372869}
\showDOI{\tempurl}


\bibitem[\protect\citeauthoryear{Green}{Green}{2021}]%
        {Green2021}
\bibfield{author}{\bibinfo{person}{Ben Green}.}
  \bibinfo{year}{2021}\natexlab{}.
\newblock \showarticletitle{{The Flaws of Policies Requiring Human Oversight of
  Government Algorithms}}.
\newblock \bibinfo{journal}{\emph{SSRN Electronic Journal}}
  (\bibinfo{year}{2021}), \bibinfo{pages}{1--42}.
\newblock
\urldef\tempurl%
\url{https://doi.org/10.2139/ssrn.3921216}
\showDOI{\tempurl}
\showeprint[arxiv]{2109.05067}


\bibitem[\protect\citeauthoryear{Green and Chen}{Green and Chen}{2019a}]%
        {Green2019a}
\bibfield{author}{\bibinfo{person}{Ben Green} {and} \bibinfo{person}{Yiling
  Chen}.} \bibinfo{year}{2019}\natexlab{a}.
\newblock \showarticletitle{{Disparate Interactions}}. In
  \bibinfo{booktitle}{\emph{Proceedings of the Conference on Fairness,
  Accountability, and Transparency}}. \bibinfo{publisher}{ACM},
  \bibinfo{address}{New York, NY, USA}, \bibinfo{pages}{90--99}.
\newblock
\showISBNx{9781450361255}
\urldef\tempurl%
\url{https://doi.org/10.1145/3287560.3287563}
\showDOI{\tempurl}


\bibitem[\protect\citeauthoryear{Green and Chen}{Green and Chen}{2019b}]%
        {Green2019}
\bibfield{author}{\bibinfo{person}{Ben Green} {and} \bibinfo{person}{Yiling
  Chen}.} \bibinfo{year}{2019}\natexlab{b}.
\newblock \showarticletitle{{The principles and limits of algorithm-in-the-loop
  decision making}}.
\newblock \bibinfo{journal}{\emph{Proceedings of the ACM on Human-Computer
  Interaction}} \bibinfo{volume}{3}, \bibinfo{number}{CSCW}
  (\bibinfo{year}{2019}).
\newblock
\showISSN{25730142}
\urldef\tempurl%
\url{https://doi.org/10.1145/3359152}
\showDOI{\tempurl}


\bibitem[\protect\citeauthoryear{Green and Chen}{Green and Chen}{2020}]%
        {Green2020}
\bibfield{author}{\bibinfo{person}{Ben Green} {and} \bibinfo{person}{Yiling
  Chen}.} \bibinfo{year}{2020}\natexlab{}.
\newblock \showarticletitle{{Algorithmic risk assessments can alter human
  decision-making processes in high-stakes government contexts}}.
\newblock  (\bibinfo{year}{2020}).
\newblock
\showeprint[arxiv]{2012.05370}
\urldef\tempurl%
\url{http://arxiv.org/abs/2012.05370}
\showURL{%
\tempurl}


\bibitem[\protect\citeauthoryear{Grgic-Hlaca, Engel, and Gummadi}{Grgic-Hlaca
  et~al\mbox{.}}{2019}]%
        {Grgic-Hlaca2019}
\bibfield{author}{\bibinfo{person}{Nina Grgic-Hlaca},
  \bibinfo{person}{Christoph Engel}, {and} \bibinfo{person}{Krishna~P.
  Gummadi}.} \bibinfo{year}{2019}\natexlab{}.
\newblock \showarticletitle{{Human decision making with machine advice: An
  experiment on bailing and jailing}}.
\newblock \bibinfo{journal}{\emph{Proceedings of the ACM on Human-Computer
  Interaction}} \bibinfo{volume}{3}, \bibinfo{number}{CSCW}
  (\bibinfo{year}{2019}).
\newblock
\showISSN{25730142}
\urldef\tempurl%
\url{https://doi.org/10.1145/3359280}
\showDOI{\tempurl}


\bibitem[\protect\citeauthoryear{Grgi{\'c}-Hla{\v{c}}a, Engel, and
  Gummadi}{Grgi{\'c}-Hla{\v{c}}a et~al\mbox{.}}{2019}]%
        {grgic2019human}
\bibfield{author}{\bibinfo{person}{Nina Grgi{\'c}-Hla{\v{c}}a},
  \bibinfo{person}{Christoph Engel}, {and} \bibinfo{person}{Krishna~P
  Gummadi}.} \bibinfo{year}{2019}\natexlab{}.
\newblock \showarticletitle{Human decision making with machine assistance: An
  experiment on bailing and jailing}.
\newblock \bibinfo{journal}{\emph{Proceedings of the ACM on Human-Computer
  Interaction}} \bibinfo{volume}{3}, \bibinfo{number}{CSCW}
  (\bibinfo{year}{2019}), \bibinfo{pages}{1--25}.
\newblock


\bibitem[\protect\citeauthoryear{Hanson, Bourgon, McGrath, Kroner, D'Amora,
  Thomas, and Tavarez}{Hanson et~al\mbox{.}}{2017}]%
        {Hanson2017}
\bibfield{author}{\bibinfo{person}{R.~Karl Hanson}, \bibinfo{person}{Guy
  Bourgon}, \bibinfo{person}{Robert~J. McGrath}, \bibinfo{person}{Daryl~G.
  Kroner}, \bibinfo{person}{David~A. D'Amora}, \bibinfo{person}{Shenique~S.
  Thomas}, {and} \bibinfo{person}{Lahiz~P. Tavarez}.}
  \bibinfo{year}{2017}\natexlab{}.
\newblock \showarticletitle{{A five-level risk and needs system: Maximizing
  assessment results in corrections through the development of a common
  language}}.
\newblock  \bibinfo{number}{January} (\bibinfo{year}{2017}).
\newblock
\urldef\tempurl%
\url{https://csgjusticecenter.org/wp-content/uploads/2017/01/A-Five-Level-Risk-and-Needs-System_Report.pdf}
\showURL{%
\tempurl}


\bibitem[\protect\citeauthoryear{Harris, Lowenkamp, and Hilton}{Harris
  et~al\mbox{.}}{2015}]%
        {Harris2015}
\bibfield{author}{\bibinfo{person}{Grant~T. Harris},
  \bibinfo{person}{Christopher~T. Lowenkamp}, {and} \bibinfo{person}{N.~Zoe
  Hilton}.} \bibinfo{year}{2015}\natexlab{}.
\newblock \showarticletitle{{Evidence for Risk Estimate Precision: Implications
  for Individual Risk Communication}}.
\newblock \bibinfo{journal}{\emph{Behavioral Sciences \& the Law}}
  \bibinfo{volume}{33}, \bibinfo{number}{1} (\bibinfo{date}{feb}
  \bibinfo{year}{2015}), \bibinfo{pages}{111--127}.
\newblock
\showISBNx{1099-0798 (Electronic)$\$r0735-3936 (Linking)}
\showISSN{07353936}
\urldef\tempurl%
\url{https://doi.org/10.1002/bsl.2158}
\showDOI{\tempurl}


\bibitem[\protect\citeauthoryear{Heilbrun, Dvoskin, Hart, and Mcniel}{Heilbrun
  et~al\mbox{.}}{1999}]%
        {Heilbrun1999}
\bibfield{author}{\bibinfo{person}{Kirk Heilbrun}, \bibinfo{person}{Joel
  Dvoskin}, \bibinfo{person}{Stephen Hart}, {and} \bibinfo{person}{Dale
  Mcniel}.} \bibinfo{year}{1999}\natexlab{}.
\newblock \showarticletitle{{Violence risk communication: Implications for
  research, policy, and practice}}.
\newblock \bibinfo{journal}{\emph{Health, Risk and Society}}
  \bibinfo{volume}{1}, \bibinfo{number}{1} (\bibinfo{year}{1999}),
  \bibinfo{pages}{91--105}.
\newblock
\showISSN{13698575}
\urldef\tempurl%
\url{https://doi.org/10.1080/13698579908407009}
\showDOI{\tempurl}


\bibitem[\protect\citeauthoryear{Hilton, Ham, Nunes, Rodrigues, Frank, and
  Seto}{Hilton et~al\mbox{.}}{2017}]%
        {Hilton2017}
\bibfield{author}{\bibinfo{person}{N.~Zoe Hilton}, \bibinfo{person}{Elke Ham},
  \bibinfo{person}{Kevin~L. Nunes}, \bibinfo{person}{Nicole~C. Rodrigues},
  \bibinfo{person}{Cairina Frank}, {and} \bibinfo{person}{Michael~C. Seto}.}
  \bibinfo{year}{2017}\natexlab{}.
\newblock \showarticletitle{{Using Graphs to Improve Violence Risk
  Communication}}.
\newblock \bibinfo{journal}{\emph{Criminal Justice and Behavior}}
  \bibinfo{volume}{44}, \bibinfo{number}{5} (\bibinfo{year}{2017}),
  \bibinfo{pages}{678--694}.
\newblock
\showISBNx{0093854816668}
\showISSN{15523594}
\urldef\tempurl%
\url{https://doi.org/10.1177/0093854816668916}
\showDOI{\tempurl}


\bibitem[\protect\citeauthoryear{Hilton, Scurich, and Helmus}{Hilton
  et~al\mbox{.}}{2015}]%
        {Hilton2015}
\bibfield{author}{\bibinfo{person}{N.~Zoe Hilton}, \bibinfo{person}{Nicholas
  Scurich}, {and} \bibinfo{person}{Leslie-Maaike Helmus}.}
  \bibinfo{year}{2015}\natexlab{}.
\newblock \showarticletitle{{Communicating the Risk of Violent and Offending
  Behavior: Review and Introduction to this Special Issue}}.
\newblock \bibinfo{journal}{\emph{Behavioral Sciences \& the Law}}
  \bibinfo{volume}{33}, \bibinfo{number}{1} (\bibinfo{date}{feb}
  \bibinfo{year}{2015}), \bibinfo{pages}{1--18}.
\newblock
\showISBNx{1099-0798 (Electronic)$\$r0735-3936 (Linking)}
\showISSN{07353936}
\urldef\tempurl%
\url{https://doi.org/10.1002/bsl.2160}
\showDOI{\tempurl}


\bibitem[\protect\citeauthoryear{Howard and Dixon}{Howard and Dixon}{2012}]%
        {howard2012construction}
\bibfield{author}{\bibinfo{person}{Philip~D Howard} {and}
  \bibinfo{person}{Louise Dixon}.} \bibinfo{year}{2012}\natexlab{}.
\newblock \showarticletitle{The construction and validation of the OASys
  Violence Predictor: Advancing violence risk assessment in the English and
  Welsh correctional services}.
\newblock \bibinfo{journal}{\emph{Criminal Justice and Behavior}}
  \bibinfo{volume}{39}, \bibinfo{number}{3} (\bibinfo{year}{2012}),
  \bibinfo{pages}{287--307}.
\newblock


\bibitem[\protect\citeauthoryear{Jahanbakhsh, Cranshaw, Counts, Lasecki, and
  Inkpen}{Jahanbakhsh et~al\mbox{.}}{2020}]%
        {jahanbakhsh2020experimental}
\bibfield{author}{\bibinfo{person}{Farnaz Jahanbakhsh}, \bibinfo{person}{Justin
  Cranshaw}, \bibinfo{person}{Scott Counts}, \bibinfo{person}{Walter~S
  Lasecki}, {and} \bibinfo{person}{Kori Inkpen}.}
  \bibinfo{year}{2020}\natexlab{}.
\newblock \showarticletitle{An Experimental Study of Bias in Platform Worker
  Ratings: The Role of Performance Quality and Gender}. In
  \bibinfo{booktitle}{\emph{Proceedings of the 2020 CHI Conference on Human
  Factors in Computing Systems}}. \bibinfo{pages}{1--13}.
\newblock


\bibitem[\protect\citeauthoryear{Jung, Pham, and Ennis}{Jung
  et~al\mbox{.}}{2013}]%
        {Jung2013}
\bibfield{author}{\bibinfo{person}{Sandy Jung}, \bibinfo{person}{Anna Pham},
  {and} \bibinfo{person}{Liam Ennis}.} \bibinfo{year}{2013}\natexlab{}.
\newblock \showarticletitle{{Measuring the disparity of categorical risk among
  various sex offender risk assessment measures}}.
\newblock \bibinfo{journal}{\emph{Journal of Forensic Psychiatry and
  Psychology}} \bibinfo{volume}{24}, \bibinfo{number}{3}
  (\bibinfo{year}{2013}), \bibinfo{pages}{353--370}.
\newblock
\showISSN{14789949}
\urldef\tempurl%
\url{https://doi.org/10.1080/14789949.2013.806567}
\showDOI{\tempurl}


\bibitem[\protect\citeauthoryear{Kleinberg, Lakkaraju, Leskovec, Ludwig, and
  Mullainathan}{Kleinberg et~al\mbox{.}}{2018}]%
        {kleinberg2018human}
\bibfield{author}{\bibinfo{person}{Jon Kleinberg}, \bibinfo{person}{Himabindu
  Lakkaraju}, \bibinfo{person}{Jure Leskovec}, \bibinfo{person}{Jens Ludwig},
  {and} \bibinfo{person}{Sendhil Mullainathan}.}
  \bibinfo{year}{2018}\natexlab{}.
\newblock \showarticletitle{Human decisions and machine predictions}.
\newblock \bibinfo{journal}{\emph{The quarterly journal of economics}}
  \bibinfo{volume}{133}, \bibinfo{number}{1} (\bibinfo{year}{2018}),
  \bibinfo{pages}{237--293}.
\newblock


\bibitem[\protect\citeauthoryear{Kr{\"o}ner, Stadtland, Eidt, and
  Nedopil}{Kr{\"o}ner et~al\mbox{.}}{2007}]%
        {kroner2007validity}
\bibfield{author}{\bibinfo{person}{Carolin Kr{\"o}ner},
  \bibinfo{person}{Cornelis Stadtland}, \bibinfo{person}{Matthias Eidt}, {and}
  \bibinfo{person}{Norbert Nedopil}.} \bibinfo{year}{2007}\natexlab{}.
\newblock \showarticletitle{The validity of the Violence Risk Appraisal Guide
  (VRAG) in predicting criminal recidivism}.
\newblock \bibinfo{journal}{\emph{Criminal Behaviour and Mental Health}}
  \bibinfo{volume}{17}, \bibinfo{number}{2} (\bibinfo{year}{2007}),
  \bibinfo{pages}{89--100}.
\newblock


\bibitem[\protect\citeauthoryear{Lee and See}{Lee and See}{2004}]%
        {Lee2004}
\bibfield{author}{\bibinfo{person}{John~D Lee} {and} \bibinfo{person}{Katrina~A
  See}.} \bibinfo{year}{2004}\natexlab{}.
\newblock \showarticletitle{{Trust in Automation: Designing for Appropriate
  Reliance}}.
\newblock \bibinfo{journal}{\emph{Human Factors: The Journal of the Human
  Factors and Ergonomics Society}} \bibinfo{volume}{46}, \bibinfo{number}{1}
  (\bibinfo{date}{jan} \bibinfo{year}{2004}), \bibinfo{pages}{50--80}.
\newblock
\showISSN{0018-7208}
\urldef\tempurl%
\url{https://doi.org/10.1518/hfes.46.1.50_30392}
\showDOI{\tempurl}


\bibitem[\protect\citeauthoryear{Lee and Selart}{Lee and Selart}{2012}]%
        {Lee2012}
\bibfield{author}{\bibinfo{person}{Wing~Shing Lee} {and}
  \bibinfo{person}{Marcus Selart}.} \bibinfo{year}{2012}\natexlab{}.
\newblock \showarticletitle{{The impact of emotions on trust decisions}}.
\newblock \bibinfo{journal}{\emph{Handbook on Psychology of Decision-Making:
  New Research}} (\bibinfo{year}{2012}), \bibinfo{pages}{235--248}.
\newblock
\showISBNx{9781621005001}


\bibitem[\protect\citeauthoryear{Lin, Jung, Goel, and Skeem}{Lin
  et~al\mbox{.}}{2020}]%
        {Lin2020}
\bibfield{author}{\bibinfo{person}{Zhiyuan~Jerry Lin}, \bibinfo{person}{Jongbin
  Jung}, \bibinfo{person}{Sharad Goel}, {and} \bibinfo{person}{Jennifer
  Skeem}.} \bibinfo{year}{2020}\natexlab{}.
\newblock \showarticletitle{{The limits of human predictions of recidivism}}.
\newblock \bibinfo{journal}{\emph{Science Advances}} \bibinfo{volume}{6},
  \bibinfo{number}{7} (\bibinfo{date}{feb} \bibinfo{year}{2020}),
  \bibinfo{pages}{1--8}.
\newblock
\showISSN{2375-2548}
\urldef\tempurl%
\url{https://doi.org/10.1126/sciadv.aaz0652}
\showDOI{\tempurl}


\bibitem[\protect\citeauthoryear{Lipkus, Samsa, and Rimer}{Lipkus
  et~al\mbox{.}}{2001}]%
        {Lipkus2001}
\bibfield{author}{\bibinfo{person}{Isaac~M. Lipkus}, \bibinfo{person}{Greg
  Samsa}, {and} \bibinfo{person}{Barbara~K. Rimer}.}
  \bibinfo{year}{2001}\natexlab{}.
\newblock \showarticletitle{{General performance on a numeracy scale among
  highly educated samples}}.
\newblock \bibinfo{journal}{\emph{Medical Decision Making}}
  \bibinfo{volume}{21}, \bibinfo{number}{1} (\bibinfo{year}{2001}),
  \bibinfo{pages}{37--44}.
\newblock
\showISSN{0272989X}
\urldef\tempurl%
\url{https://doi.org/10.1177/0272989X0102100105}
\showDOI{\tempurl}


\bibitem[\protect\citeauthoryear{Mallari, Inkpen, Johns, Tan, Ramesh, and
  Kamar}{Mallari et~al\mbox{.}}{2020}]%
        {Mallari2020}
\bibfield{author}{\bibinfo{person}{Keri Mallari}, \bibinfo{person}{Kori
  Inkpen}, \bibinfo{person}{Paul Johns}, \bibinfo{person}{Sarah Tan},
  \bibinfo{person}{Divya Ramesh}, {and} \bibinfo{person}{Ece Kamar}.}
  \bibinfo{year}{2020}\natexlab{}.
\newblock \showarticletitle{{Do I Look Like a Criminal? Examining how Race
  Presentation Impacts Human Judgement of Recidivism}}. In
  \bibinfo{booktitle}{\emph{Proceedings of the 2020 CHI Conference on Human
  Factors in Computing Systems}}. \bibinfo{publisher}{ACM},
  \bibinfo{address}{New York, NY, USA}, \bibinfo{pages}{1--13}.
\newblock
\showISBNx{9781450367080}
\urldef\tempurl%
\url{https://doi.org/10.1145/3313831.3376257}
\showDOI{\tempurl}
\showeprint[arxiv]{2002.01111}


\bibitem[\protect\citeauthoryear{McCallum, Boccaccini, and Bryson}{McCallum
  et~al\mbox{.}}{2017}]%
        {McCallum2017}
\bibfield{author}{\bibinfo{person}{Katherine~E. McCallum},
  \bibinfo{person}{Marcus~T. Boccaccini}, {and} \bibinfo{person}{Claire~N.
  Bryson}.} \bibinfo{year}{2017}\natexlab{}.
\newblock \showarticletitle{{The Influence of Risk Assessment Instrument Scores
  on Evaluators' Risk Opinions and Sexual Offender Containment
  Recommendations}}.
\newblock \bibinfo{journal}{\emph{Criminal Justice and Behavior}}
  \bibinfo{volume}{44}, \bibinfo{number}{9} (\bibinfo{year}{2017}),
  \bibinfo{pages}{1213--1235}.
\newblock
\showISSN{15523594}
\urldef\tempurl%
\url{https://doi.org/10.1177/0093854817707232}
\showDOI{\tempurl}


\bibitem[\protect\citeauthoryear{Morgan, Krueger, and King}{Morgan
  et~al\mbox{.}}{1998}]%
        {morgan1998focus}
\bibfield{author}{\bibinfo{person}{D~L Morgan}, \bibinfo{person}{R~A Krueger},
  {and} \bibinfo{person}{J~A King}.} \bibinfo{year}{1998}\natexlab{}.
\newblock \bibinfo{booktitle}{\emph{{The Focus Group Guidebook}}}.
\newblock \bibinfo{publisher}{SAGE Publications}.
\newblock
\showISBNx{9780761908180}
\urldef\tempurl%
\url{https://books.google.es/books?id=5q3k3No59OcC}
\showURL{%
\tempurl}


\bibitem[\protect\citeauthoryear{Mosier, Skitka, Heers, and Burdick}{Mosier
  et~al\mbox{.}}{1998}]%
        {Mosier1998}
\bibfield{author}{\bibinfo{person}{Kathleen~L. Mosier},
  \bibinfo{person}{Linda~J. Skitka}, \bibinfo{person}{Susan Heers}, {and}
  \bibinfo{person}{Mark Burdick}.} \bibinfo{year}{1998}\natexlab{}.
\newblock \showarticletitle{{Automation bias: Decision making and performance
  in high-tech cockpits}}.
\newblock \bibinfo{journal}{\emph{International Journal of Aviation
  Psychology}} \bibinfo{volume}{8}, \bibinfo{number}{1} (\bibinfo{year}{1998}),
  \bibinfo{pages}{47--63}.
\newblock
\showISSN{10508414}
\urldef\tempurl%
\url{https://doi.org/10.1207/s15327108ijap0801_3}
\showDOI{\tempurl}


\bibitem[\protect\citeauthoryear{Portela and Granell-canut}{Portela and
  Granell-canut}{2017}]%
        {Portela2017}
\bibfield{author}{\bibinfo{person}{Manuel Portela} {and}
  \bibinfo{person}{Carlos Granell-canut}.} \bibinfo{year}{2017}\natexlab{}.
\newblock \showarticletitle{{A new friend in our Smartphone ? Observing
  Interactions with Chatbots in the search of emotional engagement}}. In
  \bibinfo{booktitle}{\emph{Proceedings of Interacci{\'{o}}n '17}}.
\newblock
\showISBNx{9781450352291}
\urldef\tempurl%
\url{https://doi.org/10.1145/3123818.3123826}
\showDOI{\tempurl}


\bibitem[\protect\citeauthoryear{Rettenberger, M{\"o}nichweger, Buchelle,
  Schilling, and Eher}{Rettenberger et~al\mbox{.}}{2010}]%
        {rettenberger2010entwicklung}
\bibfield{author}{\bibinfo{person}{Martin Rettenberger},
  \bibinfo{person}{Michael M{\"o}nichweger}, \bibinfo{person}{Elvira Buchelle},
  \bibinfo{person}{Frank Schilling}, {and} \bibinfo{person}{Reinhard Eher}.}
  \bibinfo{year}{2010}\natexlab{}.
\newblock \showarticletitle{Entwicklung eines Screeninginstruments zur
  Vorhersage der einschl{\"a}gigen R{\"u}ckf{\"a}lligkeit von
  Gewaltstraft{\"a}tern [The development of a screening scale for the
  prediction of violent offender recidivism]}.
\newblock \bibinfo{journal}{\emph{Monatsschrift f{\"u}r Kriminologie und
  Strafrechtsreform}} \bibinfo{volume}{93}, \bibinfo{number}{5}
  (\bibinfo{year}{2010}), \bibinfo{pages}{346--360}.
\newblock


\bibitem[\protect\citeauthoryear{Sambasivan, Kapania, Highfill, Akrong,
  Paritosh, and Aroyo}{Sambasivan et~al\mbox{.}}{2021}]%
        {Sambasivan2021}
\bibfield{author}{\bibinfo{person}{Nithya Sambasivan}, \bibinfo{person}{Shivani
  Kapania}, \bibinfo{person}{Hannah Highfill}, \bibinfo{person}{Diana Akrong},
  \bibinfo{person}{Praveen Paritosh}, {and} \bibinfo{person}{Lora~M Aroyo}.}
  \bibinfo{year}{2021}\natexlab{}.
\newblock \showarticletitle{{“Everyone wants to do the model work, not the
  data work”: Data Cascades in High-Stakes AI}}. In
  \bibinfo{booktitle}{\emph{Proceedings of the 2021 CHI Conference on Human
  Factors in Computing Systems}}. \bibinfo{publisher}{ACM},
  \bibinfo{address}{New York, NY, USA}, \bibinfo{pages}{1--15}.
\newblock
\showISBNx{9781450380966}
\urldef\tempurl%
\url{https://doi.org/10.1145/3411764.3445518}
\showDOI{\tempurl}


\bibitem[\protect\citeauthoryear{Scott and Bruce}{Scott and Bruce}{1995}]%
        {Scott1995}
\bibfield{author}{\bibinfo{person}{Susanne~G. Scott} {and}
  \bibinfo{person}{Reginald~A. Bruce}.} \bibinfo{year}{1995}\natexlab{}.
\newblock \showarticletitle{{Decision-Making Style: The Development and
  Assessment of a New Measure}}.
\newblock \bibinfo{journal}{\emph{Educational and Psychological Measurement}}
  \bibinfo{volume}{55}, \bibinfo{number}{5} (\bibinfo{date}{oct}
  \bibinfo{year}{1995}), \bibinfo{pages}{818--831}.
\newblock
\showISSN{0013-1644}
\urldef\tempurl%
\url{https://doi.org/10.1177/0013164495055005017}
\showDOI{\tempurl}


\bibitem[\protect\citeauthoryear{Scurich}{Scurich}{2015}]%
        {Scurich2015}
\bibfield{author}{\bibinfo{person}{Nicholas Scurich}.}
  \bibinfo{year}{2015}\natexlab{}.
\newblock \showarticletitle{{The Differential Effect of Numeracy and Anecdotes
  on the Perceived Fallibility of Forensic Science}}.
\newblock \bibinfo{journal}{\emph{Psychiatry, Psychology and Law}}
  \bibinfo{volume}{22}, \bibinfo{number}{4} (\bibinfo{year}{2015}),
  \bibinfo{pages}{616--623}.
\newblock
\showISSN{19341687}
\urldef\tempurl%
\url{https://doi.org/10.1080/13218719.2014.965293}
\showDOI{\tempurl}


\bibitem[\protect\citeauthoryear{Scurich, Monahan, and John}{Scurich
  et~al\mbox{.}}{2012}]%
        {Scurich2012b}
\bibfield{author}{\bibinfo{person}{Nicholas Scurich}, \bibinfo{person}{John
  Monahan}, {and} \bibinfo{person}{Richard~S. John}.}
  \bibinfo{year}{2012}\natexlab{}.
\newblock \showarticletitle{{Innumeracy and unpacking: Bridging the
  nomothetic/idiographic divide in violence risk assessment}}.
\newblock \bibinfo{journal}{\emph{Law and Human Behavior}}
  \bibinfo{volume}{36}, \bibinfo{number}{6} (\bibinfo{year}{2012}),
  \bibinfo{pages}{548--554}.
\newblock
\showISSN{01477307}
\urldef\tempurl%
\url{https://doi.org/10.1037/h0093994}
\showDOI{\tempurl}


\bibitem[\protect\citeauthoryear{Selbst, Boyd, Friedler, Venkatasubramanian,
  and Vertesi}{Selbst et~al\mbox{.}}{2019}]%
        {Selbst2019}
\bibfield{author}{\bibinfo{person}{Andrew~D. Selbst}, \bibinfo{person}{Danah
  Boyd}, \bibinfo{person}{Sorelle~A. Friedler}, \bibinfo{person}{Suresh
  Venkatasubramanian}, {and} \bibinfo{person}{Janet Vertesi}.}
  \bibinfo{year}{2019}\natexlab{}.
\newblock \showarticletitle{{Fairness and abstraction in sociotechnical
  systems}}.
\newblock \bibinfo{journal}{\emph{FAT* 2019 - Proceedings of the 2019
  Conference on Fairness, Accountability, and Transparency}}
  (\bibinfo{year}{2019}), \bibinfo{pages}{59--68}.
\newblock
\showISBNx{9781450361255}
\urldef\tempurl%
\url{https://doi.org/10.1145/3287560.3287598}
\showDOI{\tempurl}


\bibitem[\protect\citeauthoryear{Singh, Grann, and Fazel}{Singh
  et~al\mbox{.}}{2011}]%
        {singh2011comparative}
\bibfield{author}{\bibinfo{person}{Jay~P Singh}, \bibinfo{person}{Martin
  Grann}, {and} \bibinfo{person}{Seena Fazel}.}
  \bibinfo{year}{2011}\natexlab{}.
\newblock \showarticletitle{A comparative study of violence risk assessment
  tools: A systematic review and metaregression analysis of 68 studies
  involving 25,980 participants}.
\newblock \bibinfo{journal}{\emph{Clinical psychology review}}
  \bibinfo{volume}{31}, \bibinfo{number}{3} (\bibinfo{year}{2011}),
  \bibinfo{pages}{499--513}.
\newblock


\bibitem[\protect\citeauthoryear{Skeem, Monahan, and Lowenkamp}{Skeem
  et~al\mbox{.}}{2016}]%
        {skeem2016gender}
\bibfield{author}{\bibinfo{person}{Jennifer Skeem}, \bibinfo{person}{John
  Monahan}, {and} \bibinfo{person}{Christopher Lowenkamp}.}
  \bibinfo{year}{2016}\natexlab{}.
\newblock \showarticletitle{Gender, risk assessment, and sanctioning: The cost
  of treating women like men.}
\newblock \bibinfo{journal}{\emph{Law and human behavior}}
  \bibinfo{volume}{40}, \bibinfo{number}{5} (\bibinfo{year}{2016}),
  \bibinfo{pages}{580}.
\newblock


\bibitem[\protect\citeauthoryear{Stevenson}{Stevenson}{2018}]%
        {stevenson2018assessing}
\bibfield{author}{\bibinfo{person}{Megan Stevenson}.}
  \bibinfo{year}{2018}\natexlab{}.
\newblock \showarticletitle{Assessing risk assessment in action}.
\newblock \bibinfo{journal}{\emph{Minnesota Law Review}}  \bibinfo{volume}{103}
  (\bibinfo{year}{2018}), \bibinfo{pages}{303}.
\newblock
\urldef\tempurl%
\url{https://heinonline.org/HOL/P?h=hein.journals/mnlr103&i=313}
\showURL{%
\tempurl}


\bibitem[\protect\citeauthoryear{Stevenson and Doleac}{Stevenson and
  Doleac}{2021}]%
        {stevenson2021algorithmic}
\bibfield{author}{\bibinfo{person}{Megan~T Stevenson} {and}
  \bibinfo{person}{Jennifer~L Doleac}.} \bibinfo{year}{2021}\natexlab{}.
\newblock \showarticletitle{Algorithmic Risk Assessment in the Hands of
  Humans}.
\newblock  (\bibinfo{year}{2021}).
\newblock
\urldef\tempurl%
\url{https://doi.org/10.2139/ssrn.3489440}
\showDOI{\tempurl}


\bibitem[\protect\citeauthoryear{Storey, Watt, and Hart}{Storey
  et~al\mbox{.}}{2015}]%
        {Storey2015}
\bibfield{author}{\bibinfo{person}{Jennifer~E. Storey},
  \bibinfo{person}{Kelly~A. Watt}, {and} \bibinfo{person}{Stephen~D. Hart}.}
  \bibinfo{year}{2015}\natexlab{}.
\newblock \showarticletitle{{An Examination of Violence Risk Communication in
  Practice Using a Structured Professional Judgment Framework}}.
\newblock \bibinfo{journal}{\emph{Behavioral Sciences \& the Law}}
  \bibinfo{volume}{33}, \bibinfo{number}{1} (\bibinfo{date}{feb}
  \bibinfo{year}{2015}), \bibinfo{pages}{39--55}.
\newblock
\showISBNx{1099-0798 (Electronic)$\$r0735-3936 (Linking)}
\showISSN{07353936}
\urldef\tempurl%
\url{https://doi.org/10.1002/bsl.2156}
\showDOI{\tempurl}


\bibitem[\protect\citeauthoryear{Tan, Adebayo, Inkpen, and Kamar}{Tan
  et~al\mbox{.}}{2018}]%
        {Tan2018a}
\bibfield{author}{\bibinfo{person}{Sarah Tan}, \bibinfo{person}{Julius
  Adebayo}, \bibinfo{person}{Kori Inkpen}, {and} \bibinfo{person}{Ece Kamar}.}
  \bibinfo{year}{2018}\natexlab{}.
\newblock \showarticletitle{{Investigating Human + Machine Complementarity for
  Recidivism Predictions}}.
\newblock  (\bibinfo{date}{aug} \bibinfo{year}{2018}).
\newblock
\showeprint[arxiv]{1808.09123}
\urldef\tempurl%
\url{http://arxiv.org/abs/1808.09123}
\showURL{%
\tempurl}


\bibitem[\protect\citeauthoryear{van Maanen, Klos, and van Dongen}{van Maanen
  et~al\mbox{.}}{2007}]%
        {VanMaanen2007}
\bibfield{author}{\bibinfo{person}{Peter-Paul van Maanen},
  \bibinfo{person}{Tomas Klos}, {and} \bibinfo{person}{Kees van Dongen}.}
  \bibinfo{year}{2007}\natexlab{}.
\newblock \showarticletitle{{Aiding Human Reliance Decision Making Using
  Computational Models of Trust}}. In \bibinfo{booktitle}{\emph{2007
  IEEE/WIC/ACM International Conferences on Web Intelligence and Intelligent
  Agent Technology - Workshops}}. \bibinfo{publisher}{IEEE},
  \bibinfo{pages}{372--376}.
\newblock
\showISBNx{0-7695-3028-1}
\urldef\tempurl%
\url{https://doi.org/10.1109/WI-IATW.2007.108}
\showDOI{\tempurl}


\bibitem[\protect\citeauthoryear{Yin, Vaughan, and Wallach}{Yin
  et~al\mbox{.}}{2019}]%
        {Yin2019}
\bibfield{author}{\bibinfo{person}{Ming Yin}, \bibinfo{person}{Jennifer~Wortman
  Vaughan}, {and} \bibinfo{person}{Hanna Wallach}.}
  \bibinfo{year}{2019}\natexlab{}.
\newblock \showarticletitle{{Understanding the effect of accuracy on trust in
  machine learning models}}.
\newblock \bibinfo{journal}{\emph{Conference on Human Factors in Computing
  Systems - Proceedings}} (\bibinfo{year}{2019}), \bibinfo{pages}{1--12}.
\newblock
\showISBNx{9781450359702}
\urldef\tempurl%
\url{https://doi.org/10.1145/3290605.3300509}
\showDOI{\tempurl}


\bibitem[\protect\citeauthoryear{Yu, Yuan, Terveen, Wu, Forlizzi, and Zhu}{Yu
  et~al\mbox{.}}{2020}]%
        {Yu2020}
\bibfield{author}{\bibinfo{person}{Bowen Yu}, \bibinfo{person}{Ye Yuan},
  \bibinfo{person}{Loren Terveen}, \bibinfo{person}{Zhiwei~Steven Wu},
  \bibinfo{person}{Jodi Forlizzi}, {and} \bibinfo{person}{Haiyi Zhu}.}
  \bibinfo{year}{2020}\natexlab{}.
\newblock \showarticletitle{{Keeping Designers in the Loop: Communicating
  Inherent Algorithmic Trade-offs Across Multiple Objectives}}. In
  \bibinfo{booktitle}{\emph{DIS'20}}. \bibinfo{pages}{1245--1257}.
\newblock
\showISBNx{9781450321389}
\urldef\tempurl%
\url{https://doi.org/10.1145/1235}
\showDOI{\tempurl}
\showeprint[arxiv]{1910.03061}


\bibitem[\protect\citeauthoryear{Zhang, Liao, Bellamy, {Vera Liao}, and
  Bellamy}{Zhang et~al\mbox{.}}{2020}]%
        {Zhang2020}
\bibfield{author}{\bibinfo{person}{Yunfeng Zhang}, \bibinfo{person}{Q.~Vera
  Liao}, \bibinfo{person}{Rachel K.E.~E. Bellamy}, \bibinfo{person}{Q. {Vera
  Liao}}, {and} \bibinfo{person}{Rachel K.E.~E. Bellamy}.}
  \bibinfo{year}{2020}\natexlab{}.
\newblock \showarticletitle{{Efect of confidence and explanation on accuracy
  and trust calibration in AI-assisted decision making}}.
\newblock \bibinfo{journal}{\emph{FAT* 2020 - Proceedings of the 2020
  Conference on Fairness, Accountability, and Transparency}}
  (\bibinfo{date}{jan} \bibinfo{year}{2020}), \bibinfo{pages}{295--305}.
\newblock
\showISBNx{9781450369367}
\urldef\tempurl%
\url{https://doi.org/10.1145/3351095.3372852}
\showDOI{\tempurl}
\showeprint[arxiv]{2001.02114}


\bibitem[\protect\citeauthoryear{{Zoe Hilton}, Carter, Harris, and Sharpe}{{Zoe
  Hilton} et~al\mbox{.}}{2008}]%
        {ZoeHilton2008}
\bibfield{author}{\bibinfo{person}{N. {Zoe Hilton}}, \bibinfo{person}{Angela~M.
  Carter}, \bibinfo{person}{Grant~T. Harris}, {and} \bibinfo{person}{Amilynn
  J.~B. Sharpe}.} \bibinfo{year}{2008}\natexlab{}.
\newblock \showarticletitle{{Does Using Nonnumerical Terms to Describe Risk Aid
  Violence Risk Communication?}}
\newblock \bibinfo{journal}{\emph{Journal of Interpersonal Violence}}
  \bibinfo{volume}{23}, \bibinfo{number}{2} (\bibinfo{year}{2008}),
  \bibinfo{pages}{171--188}.
\newblock
\showISBNx{0886260507}
\showISSN{0886-2605}
\urldef\tempurl%
\url{https://doi.org/10.1177/0886260507309337}
\showDOI{\tempurl}


\bibitem[\protect\citeauthoryear{Zuiderwijk, Chen, and Salem}{Zuiderwijk
  et~al\mbox{.}}{2021}]%
        {Zuiderwijk2021}
\bibfield{author}{\bibinfo{person}{Anneke Zuiderwijk}, \bibinfo{person}{Yu~Che
  Chen}, {and} \bibinfo{person}{Fadi Salem}.} \bibinfo{year}{2021}\natexlab{}.
\newblock \showarticletitle{{Implications of the use of artificial intelligence
  in public governance: A systematic literature review and a research agenda}}.
\newblock \bibinfo{journal}{\emph{Government Information Quarterly}}
  \bibinfo{number}{May 2020} (\bibinfo{year}{2021}), \bibinfo{pages}{101577}.
\newblock
\showISSN{0740624X}
\urldef\tempurl%
\url{https://doi.org/10.1016/j.giq.2021.101577}
\showDOI{\tempurl}


\end{thebibliography}

\appendix

\noindent{\Large\textbf{SUPPLEMENTARY MATERIAL}}

\section{Additional information about our approach}

\subsection{Experimental groups}

The number of participants in each experimental group is shown in Table~\ref{tab:experimentalgroups}.

\begin{table}[h]
\small
    \begin{tabular}{l|cl|cl|cl}
         Type $\rightarrow$ 
         & \multicolumn{2}{c|}{\textbf{Crowdsourced}}
         & \multicolumn{2}{c|}{\textbf{Crowdsourced}}
         & \multicolumn{2}{c}{\textbf{Targeted}} \\
         Group $\downarrow$ 
         & {N} & {Round 1}
         & {N} & {Round 2} 
         & {N} &  ~ \\
         \hline
         Control 
         & 48 & Abs. scale
         & 17 & Abs. scale 
         & - & - 
         \\
         G1 
         & 100 & Abs. scale / categorical
         &  66 & Abs. scale / cat. and num. 
         & 36 &  Abs. scale / cat. and num.
         \\
         G2 
         & 99 & Abs. scale / cat. and num.
         &  63 & Rel. scale / cat. and num.
         & 18 & Rel. scale / cat. and num.
         \\
         \hline 
         Total
             & 247 &
             & 146 & 
             & 54 \\
    \end{tabular}
    \caption{Characteristics of the experimental groups.
    The control groups received no machine predictions.
    The treatment groups received machine predictions.
    G1 used an absolute scale indicating a probability (0\% to 100\%); 
    G2 used a relative scale indicating a score (0 to 10).}
    \label{tab:experimentalgroups}
\end{table}

\subsection{Designing the algorithm}

We use logistic regression to predict violent recidivism. 
The features given as input are the 23 items that determine the REVI score in \RisCanvi, plus three demographic features (age, gender, and nationality).
The evaluation was done by $k$-fold cross-validation, i.e., dividing the data into $k$ parts, training on $k-1$ parts and evaluating on the remaining part.
The accuracy of the model is 0.76 in terms of AUC-ROC which is the average result over the $k$ runs.
Finally, the logistic regression estimates were calibrated, which means that they were transformed to correspond to an estimate of the probability of the outcome.

\subsection{Datasets} \label{ann:casesel}

An original dataset with 597 cases was used for creating the algorithm as described above. 
The dataset is anonymized and shared through a formal collaboration agreement between our university and the Department of Justice of \Catalonia.
This agreement indicates that no personal data is shared with the university.
%

\subsubsection{Semi-synthetic case pool (90 cases)}

Although the original dataset did not include personal information, we wanted to make sure that participants never had access to the features of one person.
Hence, we created 90 semi-synthetic cases by doing a cross-over of features within a group of similar cases.
Each group of cases was selected so that the difference in computed \RisCanvi between the highest and lowest risk was at most 0.1.
The generated case differs by a minimum of one and a maximum of three features from any case in the group, and has a \RisCanvi risk level within the same risk range of the cases in each group.
%
%
%
%
A preliminary experiment with 31 crowdsourced participants, in which no machine assistance was shown, was used to estimate the difficulty of human risk assessment (i.e., how distant was the prediction from the ground truth) for each case.

\subsubsection{Case selection (11 cases)}

From 90 semi-synthetic cases, 11 cases were selected to be evaluated by participants. 
This selection was done by sampling 8 non-recidivists and 3 recidivists to have a recidivism rate close to what is observed in \Catalonia.
To perform this sampling, we stratified the cases by human difficulty and machine difficulty into three groups (easy, medium, hard).
``Difficulty'' means how far, on average, is the prediction from the ground truth.
This yields nine classes of difficulty (e.g., ``easy'' for humans and ``hard'' for the model) from which we sampled the 11 cases.
%
As we explained in Section \ref{sec:methods}, we exchanged two cases to increase the general AUC-ROC of the entire dataset.

\begin{figure}
  \includegraphics[width=\textwidth]{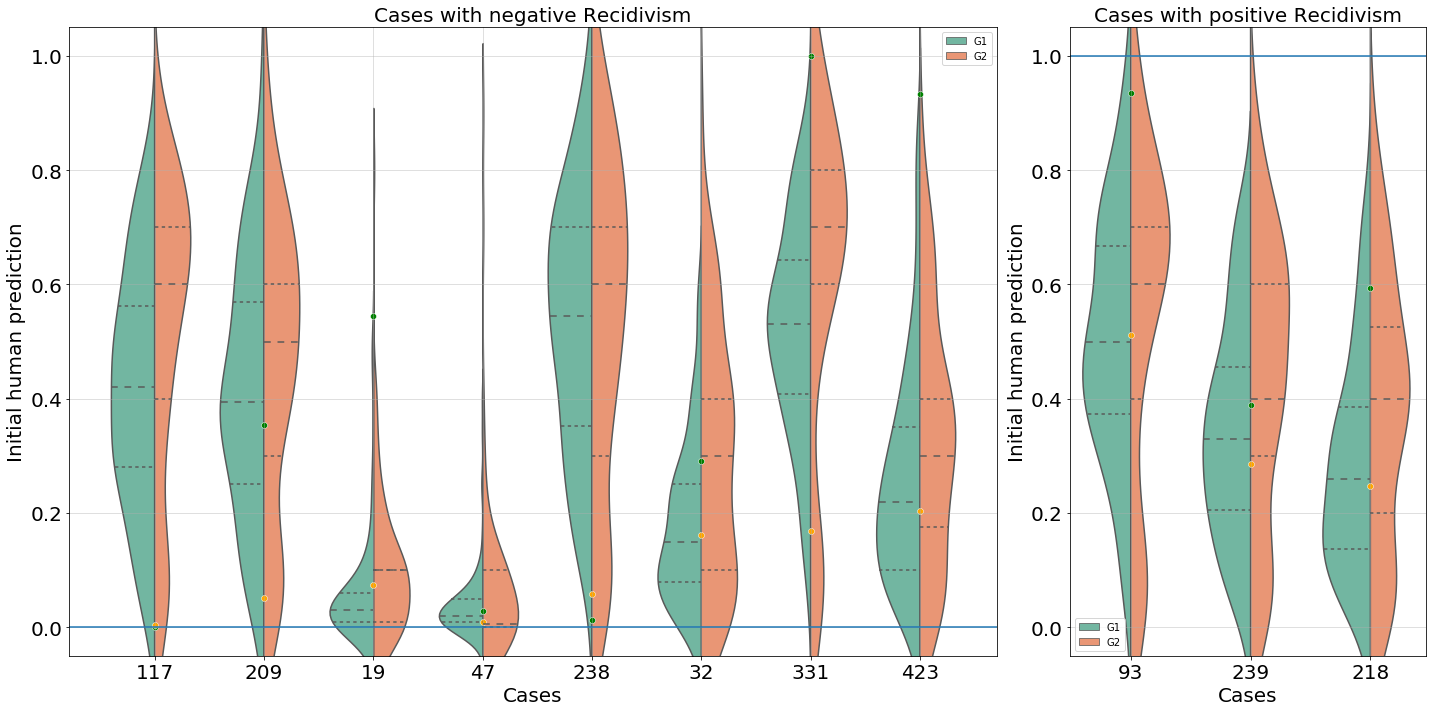}
   \caption{Comparison of human predictions distribution on 11 experimental cases, on absolute scale setting (G1, in green) and relative scale setting (G2, in orange).
   The yellow dot indicates the machine prediction for each case.
   The first 8 cases are non-recidivists (left) and the last 3 are recidivists (right). We remark that all 11 cases were shown in random order to each participant.
   }
  \label{fig:casespredictions}
\end{figure}

The resulting 11 cases are depicted in Figure \ref{fig:casespredictions}, where cases are grouped by ground truth and their risks are predicted by crowdsourced (R2) and targeted studies combined.
%
It can be noticed that the accuracy of human predictions differs for different cases, and that in some cases, answers are more spread. 

\section{Surveys}

\subsection{Level of automation survey}
\label{ann:automationlevel}

This survey was based on the levels of automation proposed by Cummings \cite{Cummings2004}.
Cummings proposed ten levels going from ``the computer decides everything and acts autonomously, ignoring the human'' (level 10) to ``the computer offers no assistance: a human must take all decisions and actions'' (level 1).
We reduced the ten levels to five, to make it more understandable and easier to answer for participants. 

\noindent\textbf{Question: Would you use a computer system, developed at a university, and based on statistics, to predict the level of risk of violent criminal recidivism?}
\begin{itemize}
    \item Level 1: No, the expert should decide the risk level by himself/herself
    \item Level 2: Only if the computational system suggests some options, but it is the expert who defines the risk level
    \item Level 3: Only if the computational system suggests one option, but the expert can provide an alternative option
    \item Level 4: Only if the computational system suggests one option and the expert can decide to take it or not
    \item Level 5: Yes, the computational system should decide the risk level by itself
\end{itemize}

\subsection{Questions for the focus groups}\label{ann:focusgroup}

Questions were used to stimulate the discussion, but we invited participants to comment on any aspect of the experiment.

\begin{itemize}
 \item \textbf{Q1:} What is your general opinion on Risk Assessment Instruments [explain to participants] in criminal justice settings? 
    \item \textbf{Q2:} How do you think the machine prediction in this study works? Could you explain it? 
    \item \textbf{Q3:} Do you think that one of these two scales [show them to participants] would be better than the other? Why?
         \item    \textbf{Q4:} From the list of case characteristics [show to participants], which were the ones that helped you the most to make a decision about the risk of recidivism?
         \item    \textbf{Q5:} Explain, why do you think that these features can help define the prediction of these cases?
         \item    \textbf{Q6:} What does a 10\% risk mean to you in the context of this study? 
         \item    \textbf{Q7:} What does a 2 over 10 risk mean for you in the context of the study?
    \item \textbf{Q8:} Despite an improvement in the accuracy, participants tended to rely less on the machine prediction after the experiment, why do you think that it happens? What was your experience?
        \item     \textbf{Q9:} Suppose that you can decide to use an algorithm-supported decision making system in this context. What would be the advantages and disadvantages of it?
\end{itemize}

\subsection{Assessment Interface}
\label{ann:surveytemplate}

\begin{figure}[ht]
\begin{minipage}[b]{0.45\linewidth}
  \includegraphics[width=\linewidth]{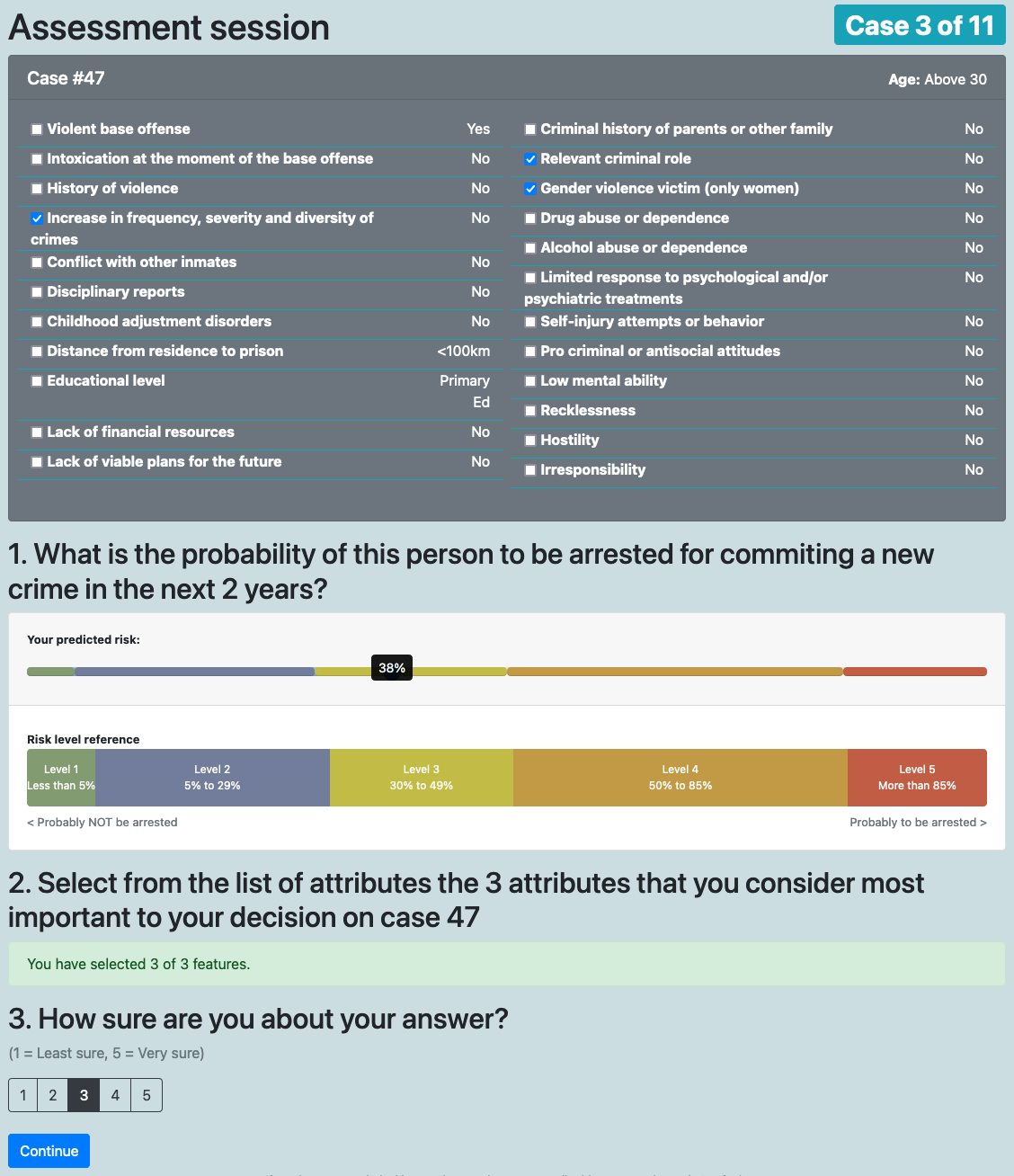}
  \caption{Evaluation session first page: User prediction and selection of risk factors.}
  \label{fig:survey_1}
  \Description{First page of evaluation.}
 \end{minipage}
 \begin{minipage}[b]{0.45\linewidth}
  \includegraphics[width=\linewidth]{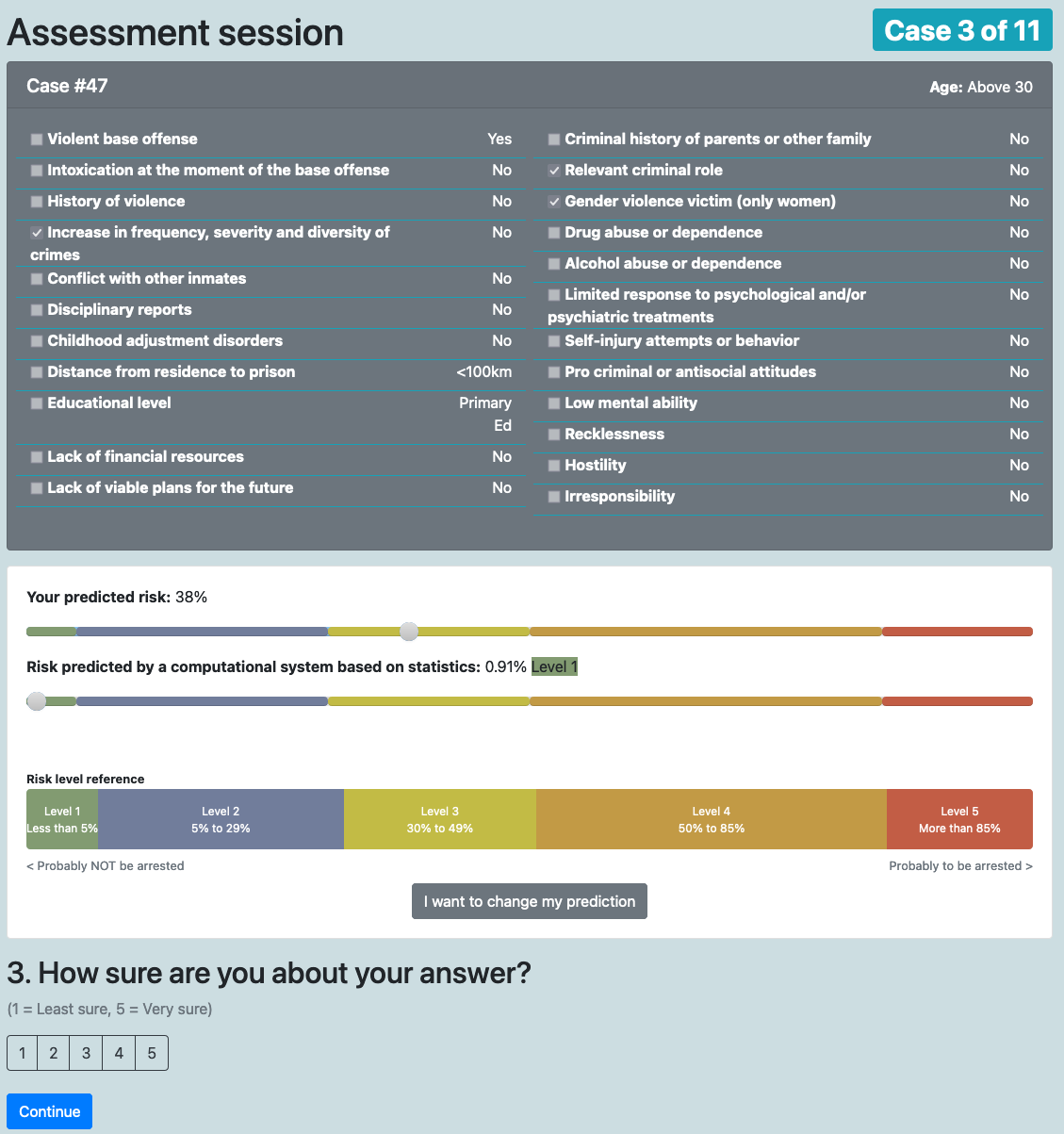}
  \caption{Evaluation session second page: Algorithm prediction and user confirmation.}
  \label{fig:survey_2}
  \Description{Second page of evaluation.}
 \end{minipage}
\end{figure}

Figure~\ref{fig:survey_1} shows the first page of each evaluation, which is the same for the control and treatment groups. Participants can see each of the risk factors with their variables and select those 3 that they believe are more important to define their prediction. They are asked to define their probability moving the marker on the bar, having as a reference the five levels of risk that depend on the type of scale used. Before moving to the next page they have to assign a value to their confidence in their answer.
Figure~\ref{fig:survey_2} shows the second page with the machine assistance, which only the treatment groups see. It shows the algorithm prediction in a similar bar/scale, compared to the participant's prediction. Then, the participant has the possibility to change their own prediction and provide a confidence score. 

\begin{figure}
  \includegraphics[width=0.5\textwidth]{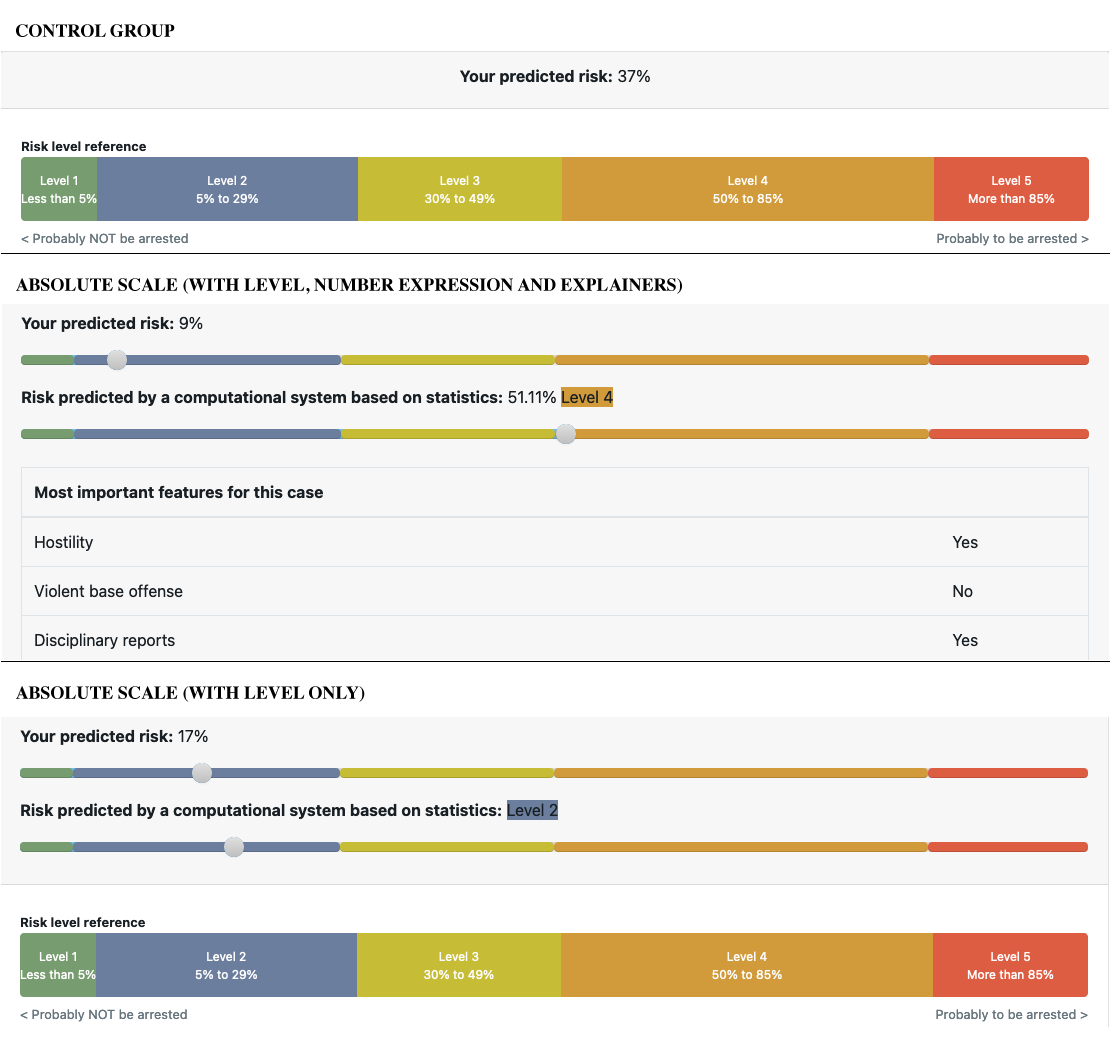}
  \caption{Alternative treatment groups in crowdsourced R1 study}
  \label{fig:scales_alt}
\Description{A woman and a girl in white dresses sit in an open car.}
\end{figure}

Figure~\ref{fig:scales_alt} shows the alternatives in the crowdsourced R1 study. The control group only sees their own prediction without any feedback. The G1 (bottom) is able to see their prediction compared with the algorithm's prediction, while G2 (middle) is able to see also the most important features that defines the algorithm's prediction (explainers).

\clearpage

\section{Additional results}
\label{ann:additionalresults}

\subsection{Accuracy results per sub-group}
\label{ann:accuracy-per-subgroup}

\begin{table*}[t]
  \resizebox{0.95\textwidth}{!}{%
  \begin{tabular}{l|ccccccc}
    \toprule
      Groups $\rightarrow$  & Control & \multicolumn{2}{c}{R1G1}  & 
      \multicolumn{2}{c}{R1G2 | R2G1 | TG1} &  
      \multicolumn{2}{c}{R2G2 | TG2}\\
      Treatment $\rightarrow$ & Absolute & \multicolumn{2}{c}{Absolute \& non-numerical}  & \multicolumn{2}{c}{Absolute \& percentage} &  \multicolumn{2}{c}{Relative \& score}\\
      Participant type $\downarrow$ 
      &  & Before & After & Before & After & Before & After \\
    \midrule
    Crowdsourced R1 - N=247 & 0.599 \textcolor{gray}{(0.277)} & 0.598 \textcolor{gray}{(0.273)} & 0.621 \textcolor{gray}{(0.271)}  & 0.576 \textcolor{gray}{(0.291)} & 0.607 \textcolor{gray}{(0.280)} & - &  - \\
    Crowdsourced R2 - N=146 & 0.652 \textcolor{gray}{(0.262)} & - & - & 0.582 \textcolor{gray}{(0.265)} & 0.665 \textcolor{gray}{(0.240)} & 0.606 \textcolor{gray}{(0.271)} & 0.622 \textcolor{gray}{(0.269)}   \\
    \midrule
    Data science (students) - N=14 & - & - & - & 0.652 \textcolor{gray}{(0.229)} & 0.690 \textcolor{gray}{(0.203)} & 0.589 \textcolor{gray}{(0.269)} & 0.667 \textcolor{gray}{(0.266)}  \\
    Data science (professionals) - N=11 & - & - & - & 0.736 \textcolor{gray}{(0.219)} & 0.729 \textcolor{gray}{(0.218)} & 0.667 \textcolor{gray}{(0.269)} & 0.719 \textcolor{gray}{(0.267)}  \\
    Domain experts (students) - N=4 & - & - & -
    & 0.771 \textcolor{gray}{(0.211)} & 0.833 \textcolor{gray}{(0.169)}  & 0.750 \textcolor{gray}{(0.264)} & 0.729 \textcolor{gray}{(0.264)}  \\
    Domain experts (professionals) - N=25 & - & - & -
    & 0.636 \textcolor{gray}{(0.236)} & 0.773 \textcolor{gray}{(0.191)}  & 0.637 \textcolor{gray}{(0.270)} & 0.673 \textcolor{gray}{(0.267)} \\
    \bottomrule
  \end{tabular}}
  \caption{AUC-ROC (and Brier score in \textcolor{gray}{gray}) before and after algorithmic support by experimental group, type of risk scale and numerical expression.}
  \label{tab:results1}
\end{table*}

In Table \ref{tab:results1} we present the results from different groups, including targeted groups of data scientists and domain experts, separating students and professionals.
In the table we also include the Brier Score (lower is better), which is consistent with the AUC-ROC results (higher is better).

In addition, in Table \ref{tab:ttests} we show results from a t-test to show differences between crowdsourced and targeted groups.

\begin{table}[ht] 
\centering
\begin{tabular}{lcrr}
  \hline
 Group 1 & Relation & Group 2 & p-value \\ 
  \hline
Crowd R1- Control       & < & Targeted - G1 Before & 0.018 \\ 
Crowd R1- Control     & < & Targeted - G1 After & 0.002 \\ 
Crowd R1- Control     & < & Targeted - G2 After & 0.016 \\ 
Crowd R1 - G1 Before  & < & Targeted - G1 Before & 0.002 \\ 
Crowd R1 - G1 Before  & < & Targeted - G1 After & 0.002 \\ 
Crowd R1 - G1 Before  & < & Targeted - G2 After & 0.006 \\ 
Crowd R1 - G2 Before  & < & Targeted - G1 Before & 0.004 \\ 
Crowd R1 - G2 Before  & < & Crowd R1 - G1 After & 0.020 \\ 
Crowd R1 - G2 Before  & < & Crowd R1 - G2 After & 0.092 \\ 
Crowd R2- Control    & \textcolor{red}{>} & Crowd R2 - G1 Before & 0.082 \\ 
Crowd R2- Control    & < & Targeted - G1 After & 0.010 \\ 
Crowd R2 - G1 Before & < & Targeted - G1 Before & 0.004 \\ 
Crowd R2 - G1 Before & < & Crowd R2 - G1 After & 0.012 \\ 
Crowd R2 - G1 Before & < & Targeted - G1 After & 0.002 \\ 
Crowd R2 - G1 Before & < & Targeted - G2 After & 0.018 \\ 
Crowd R2 - G2 Before & < & Targeted - G1 Before & 0.042 \\ 
Crowd R2 - G2 Before & < & Crowd R2 - G1 After & 0.044 \\ 
Crowd R2 - G2 Before & < & Targeted - G1 After & 0.002 \\ 
Crowd R2 - G2 Before & < & Targeted - G2 After & 0.034 \\ 
Targeted - G1 Before & \textcolor{red}{>} & Crowd R2 - G2 After & 0.088 \\ 
Targeted - G1 Before & < & Targeted - G1 After & 0.026 \\ 
Targeted - G2 Before & < & Targeted - G1 After & 0.004 \\ 
Crowd R2 - G1 After  & < & Targeted - G1 After & 0.020 \\ 
Crowd R2 - G2 After  & < & Targeted - G1 After & 0.002 \\ 
Targeted - G1 After  & \textcolor{red}{>} & Targeted - G2 After & 0.058 \\ 
   \hline
   
\end{tabular}
\caption{Permutation t-tests for differences between AUC-ROC of subgroups, showing only differences that are significant at $p < 0.1$.
In most cases the relation is that crowdsourced groups are less accurate than targeted groups, and predictions before machine assistance are less accurate than predictions after machine assistance. Cases where this relationship is inverted are more rare. The relationship signs ($<,>$) indicate which group has a higher AUC.}\label{tab:ttests}
\end{table}

\subsection{Preferred level of automation and experience}
\label{ann:preferred-level-of-automation}

In addition to the level of automation question, we asked participants about their previous experience with Risk Assessment Instruments (RAI) on a scale from 1 (This is the first time I heard or read about risk assessment instruments) to 5 (I have used this kind of tools more than one time).
In Table \ref{tab:rai} we can see that targeted participants report more previous experience than crowdsourced participants. 
We can also see that the preferred level of automation is in general lower at the end of the experiment than at the beginning, in all cases except for the control groups.

\begin{table}[ht]
  \begin{tabular}{lccc}
    \toprule
     Experimental Groups & Experience  & \multicolumn{2}{c}{Preferred level of automation}\\
                         & (1=least, 5=most) & \multicolumn{2}{c}{(1=no automation, 5=fully automated)} \\
    & & Start & End\\
    \midrule
    Crowd R1 - Control & 1.62 $\pm$ 0.76 & 2.88 $\pm$ 1.02  & 2.96 $\pm$ 1.03 \\
    Crowd R1 - G1 & 1.79 $\pm$ 0.99 & 3.08 $\pm$ 1.07 & 3.01 $\pm$ 1.05 \\
    Crowd R1 - G2 & 1.70 $\pm$ 0.82 & 3.01 $\pm$ 1.08 & 2.87 $\pm$ 1.10 \\
    Crowd R2 - Control & 1.29 $\pm$ 0.47 & 2.82 $\pm$ 1.07 & 2.59 $\pm$ 1.00  \\
    Crowd R2 - G1 & 1.48 $\pm$ 0.67 & 2.61 $\pm$ 1.03 & 2.48 $\pm$ 1.06 \\
    Crowd R2 - G2 & 1.70 $\pm$ 0.85 & 3.00 $\pm$ 1.03 &  2.60 $\pm$ 0.91 \\
    Targeted G1 & 3.02 $\pm$ 1.43 &   2.75 $\pm$ 0.90 & 2.68 $\pm$ 0.84 \\
    Targeted G2 & 3.16 $\pm$ 1.44 &   3.11 $\pm$ 0.83 & 3.05 $\pm$ 0.87 \\
    \bottomrule
  \end{tabular}
  \caption{Users' experience with RAIs (scale of 1 to 5) and preferred level of automation (scale of 1 to 5,  as described in \S\ref{ann:automationlevel}) at the start and at the end of the study, including standard deviation values.}\label{tab:rai}
\end{table}
    
\subsection{Self-reported confidence}
\label{ann:self-reported-confidence}

We observe that self-reported confidence is stable across all subgroups.
For each case evaluation, participants had to answer their level of confidence on a likert scale (1 to 5).
After seeing the algorithm prediction and with the opportunity to change their prediction or not, they are asked about their confidence level again. 
Results are shown in Table~\ref{tab:confidence}.

\begin{table}[t]
\begin{minipage}{0.5\linewidth}
     \resizebox{\linewidth}{!}{%
     \begin{tabular}{ll|cc}
    \toprule
      Study & Group & Before & After \\
    \midrule
    Crowd (R2) & G1 & 3.89 $\pm$ 0.89 & 3.82  $\pm$ 0.98 \\
    Crowd (R2) & G2 & 3.90  $\pm$ 0.77 & 3.82  $\pm$ 0.95 \\
    Targeted & G1 & 3.48  $\pm$ 0.71 & 3.45  $\pm$ 0.85 \\
    Targeted & G2 & 3.57  $\pm$ 0.81 & 3.52  $\pm$ 0.97 \\
    \midrule
    Targeted & Domain Experts & 3.47 $\pm$ 0.73 & 3.52  $\pm$ 0.82 \\
    Targeted & Data Scientists & 3.55 $\pm$ 0.76 & 3.43  $\pm$ 1.04 \\
    \bottomrule
  \end{tabular}

    }\caption{Average self-reported confidence by subgroup, with standard deviation values, before and after seeing the machine prediction.
    1=least confident, 5=most confident.}\label{tab:confidence}
\end{minipage}%
\hspace{5mm}%
\begin{minipage}{0.45\linewidth}
    \resizebox{\linewidth}{!}{%
    
\begin{tabular}{lll}
\toprule Group 1 & Group 2 & p-value \\\midrule
Crowd R2 - G1 Initial & Targeted - G1 Initial & $<$0.001**** \\ 
Crowd R2 - G2 Initial & Targeted - G2 Initial & $<$0.001**** \\ 
Crowd R2 - G2 Initial & Targeted - G1 Initial & $<$0.001**** \\ 
Crowd R2 - G2 Initial & Targeted - G2 Initial & $<$0.001**** \\ 
Crowd R2 - G2 Initial & Targeted - G2 Final & $<$0.001****\\ 
Crowd R2 - G1 Initial & Targeted - G1 Final	& $<$0.001****\\ 
Crowd R2 - G1 Initial & Targeted - G2 Final	& $<$0.001****\\ 
Crowd R2 - G2 Initial & Targeted - G1 Final	& $<$0.001****\\
\toprule
\end{tabular}

    }\caption{Paired t-test for self-reported confidence, only only p <0.001is shown.}\label{tab:confidence_ttest}
    
\end{minipage}
 \end{table}

\subsection{Decision-making style and emotional state results}
\label{ann:gdms-vas}
During the survey we included two surveys to measure the current emotional state (e.g., joyful, sad, anger) and decision making style (e.g., rational, intuitive). We used VAS ~\cite{Portela2017} for the emotional state, and GDMS ~\cite{Scott1995} for the decision making style.

Results reflected in Table \ref{tab:dgms_vas} are similar across subgroups.
Common emotional states reported are joyful, relaxed, and energized.
In general, intuitive and spontaneous decision making appears with higher levels than rational. 
%
\begin{table}[t]
  \caption{DGMS and VAS results}
  \label{tab:dgms_vas}
  \begin{tabular}{l|ccc|ccc|cc}
    \toprule
    Survey  &  \multicolumn{3}{c}{Crowd. (R1)} & \multicolumn{3}{c}{Crowd. (R2)} & \multicolumn{2}{c}{Targeted} \\
            & GC & G1 & G2 & GC & G1 & G2 & G1 & G2 \\
    \midrule
    VAS \\
        Joy & 3.27 & 3.51 & 3.45 & 3.59 & 3.55 & 3.92 & 3.89 & 3.72 \\
        Sad & 2.29 & 2.08 & 2.24 & 2.29 & 2.17 & 1.63 & 1.97 & 2.11 \\
        Angry & 1.60 & 1.65 & 1.59 & 1.71 & 1.83 & 1.46 & 1.56 & 1.22 \\
        Surprise & 1.75 & 1.77 & 1.86 & 1.71 & 1.86  & 2.13 & 1.72 & 2.06 \\
        Relax & 3.79 & 3.59 & 3.74 & 3.47 & 3.55 & 3.94 & 3.33 & 3.50 \\
        Energy & 2.94 & 3.05 & 2.80 & 2.88 & 3.09 & 3.32 & 3.31 & 3.00 \\
    \midrule
    GDMS   \\
        Rational  & 0.61 & 0.61 & 0.64 & 0.63 & 0.60 & 0.62 & 0.58 & 0.59 \\
        Intuitive  & 0.72 & 0.72 & 0.71 & 0.72 & 0.75 & 0.74 & 0.71 & 0.68 \\
        Dependent  & 0.59 & 0.59 & 0.64 & 0.55 & 0.58 & 0.59 & 0.57 & 0.53 \\
        Avoidant  & 0.59 & 0.59 & 0.64 & 0.61 & 0.59 & 0.61 & 0.57 & 0.54 \\
        Spontaneous  & 0.68 & 0.69 & 0.71 & 0.68 & 0.69 & 0.71 & 0.71 & 0.70 \\
   
    \bottomrule
  \end{tabular}
\end{table}

\subsection{\textit{RiskEval} items considered as most important}
\label{ann:items-considered-most-important}

\begin{figure}[t]
\centering
\includegraphics[width=\textwidth]{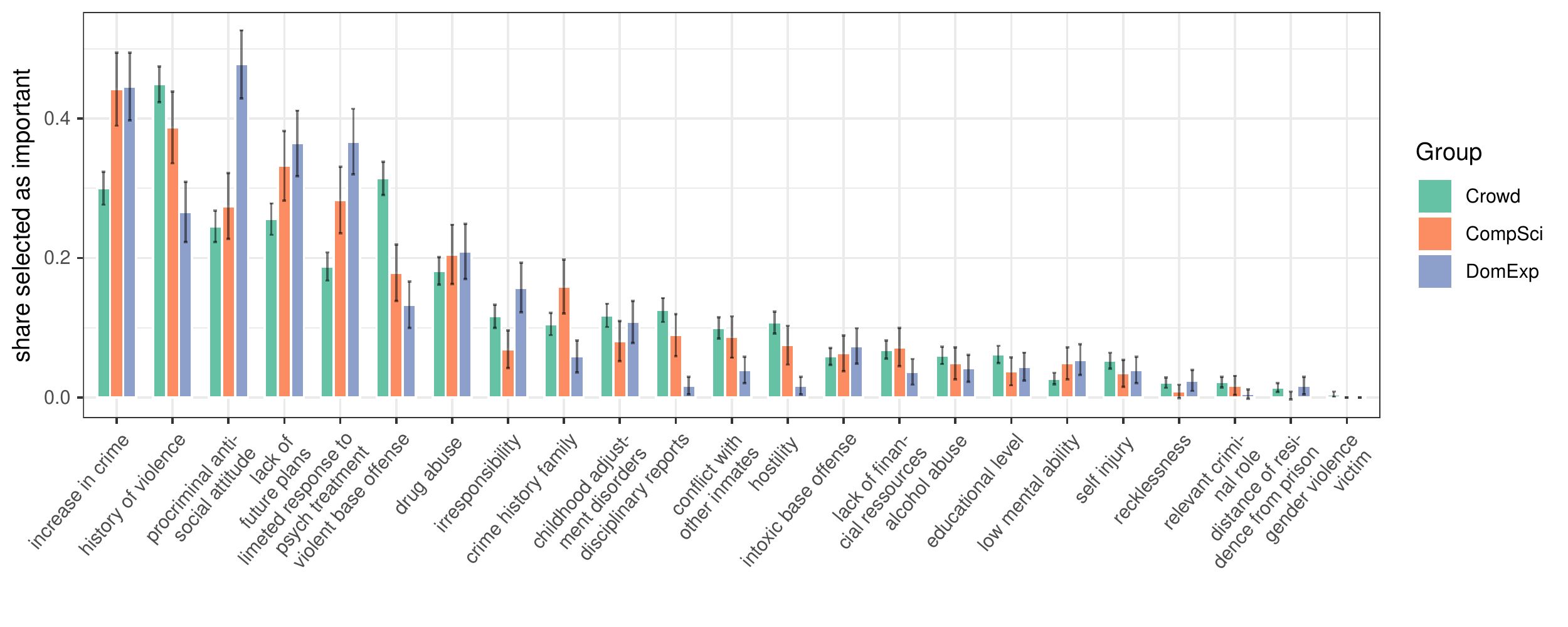}
\captionof{figure}{Amount of times that each item was selected as important to the decision, by participants with different backgrounds (crowdsourced, data science and domain experts)}
\label{fig:items_important}
\end{figure}

As explained in subsection \ref{importanceofriskitems}, the top items (features) selected as important for most participants tend to be the same (Figure~\ref{fig:items_important}).
Targeted groups of data scientists and domain experts selected the same top five items, albeit in a different ordering, and their top items overlap to some extent with those of crowdsourced participants (Table~\ref{tab:top5_items}).
%

\begin{table}[t]
\centering

\begin{tabular}{lp{4cm}p{4cm}p{4cm}}
  \hline
  & Crowdsourced
  & Targeted: Data Science 
  & Targeted: Domain Experts \\ 
  \hline
$1\textsuperscript{st}$ & History of violence & Increase in frequency, severity and diversity of crimes & Pro criminal or antisocial attitude \\\hline 
$2\textsuperscript{nd}$ & Violent base offense & History of violence & Increase in frequency, severity and diversity of crimes \\\hline 
$3\textsuperscript{rd}$ & Increase in frequency, severity and diversity of crimes & Lack of viable plans for the future & Limited response to psychological/psychiatric treatment \\\hline 
$4\textsuperscript{th}$ & Lack of viable plans for the future & Limited response to psychological/psychiatric treatment & Lack of viable plans for the future \\\hline 
$5\textsuperscript{th}$ & Pro criminal or antisocial attitude & Pro criminal or antisocial attitude & History of violence \\ 
   \hline
\end{tabular}
\caption{Top 5 items (features) listed as most important for making decisions by different background}\label{tab:top5_items}
\end{table}

\end{document}